\documentclass[superscriptaddress,reprint,aps]{revtex4-2}
\usepackage{amsmath}
\usepackage{amssymb}
\usepackage{graphicx}
\usepackage{srcltx}
\usepackage{bm}
\usepackage{color}
\usepackage{ifthen}
\usepackage{fancybox}
\usepackage{datetime}
\usepackage{soul} % for using the command st

\def\ignore#1{}
\def\fignore#1{}

\def\figbox#1{\Blue\fbox{#1} \Black}
\def\figbox#1{}

\def\be{\begin{equation}}
\def\ee{\end{equation}}
\def\bea{\begin{eqnarray}}
\def\eea{\end{eqnarray}}

\def\nn{\nonumber\\}

\def\Blue{}

\def\Black{}

\def\ignore#1{}

\newcommand{\ba}{\begin{align}}
\newcommand{\ea}{\end{align}}

\begin{document}
\title{
	Photon Transport in   a Gas of Two-Level Atoms: Unveiling Quantum Light Creation
}
\author{Leonid Yatsenko}
\affiliation{ Institute of Physics, National Academy of Science of Ukraine,  Nauki Avenue 46, Kyiv 03028, Ukraine}

\author{ Martin Cordier}
\author{ Lucas Pache }
\author{ Max Schemmer   }
\altaffiliation{New address: Istituto Nazionale di Ottica, Consiglio Nazionale delle Ricerche, 50019 Sesto Fiorentino, Italy}
\author{ Philipp Schneeweiss  }
 \author{ J\"{u}rgen Volz}
 \author{  Arno Rauschenbeutel}
 \affiliation{Department of Physics, Humboldt-Universit\"{a}t zu Berlin, Unter den Linden 6, 10099 Berlin, Germany }
\begin{abstract}

We present a theoretical analysis of nearly monochromatic light propagation through a gas of two-level atoms   using the Heisenberg-Langevin equation method. Our focus is on the evolution of the  photon annihilation operator   and its impact on the second-order correlation function, \( g^{(2)}(\tau) \), with particular emphasis on photon antibunching behavior. The model accounts for both open and closed atomic system approximations, including Doppler broadening and the influence of pump field detuning. We derive expressions that reproduce known results from scattering theory and extend the analysis to complex systems, such as inhomogeneously broadened media. The theoretical predictions are compared with experimental data from a waveguide QED platform, which show good agreement and thereby demonstrate the power of our approach. Future work will explore extensions to even more complex systems and other quantum light characteristics for practical applications.
\end{abstract}

\maketitle

\section{Introduction}
\label{intro}

The propagation of light through atomic media has long been a topic of great interest in quantum optics and atomic physics \cite{Ham10}. When nearly monochromatic light interacts with an absorbing gas, a rich variety of quantum phenomena emerges, driven by the intricate interplay between the light field and the atomic coherence it induces \cite{Alle87, Fle05}. This interaction is fundamental to understanding a wide range of applications, from the study of fundamental quantum effects \cite{Har06}  to the creation of novel quantum light sources \cite{Kolo07}.

The behavior of light in resonantly absorbing media can be profoundly altered by quantum effects, leading to phenomena such as electromagnetically induced transparency \cite{Fle05}, photon blockade \cite{Bir05}, and the generation of non-classical states of light   \cite{Wal08}. Traditional approaches to studying these effects often rely on  perturbative scattering methods \cite{Dal83,Coh98}. However, these methods may fall short in capturing the full quantum nature of the light-matter interaction, particularly in more complex systems \cite{Har06}.

In this work, we present a theoretical analysis of the quantum behavior of nearly monochromatic light as it propagates through a gaseous medium composed  of two-level atoms. Our approach utilizes the Heisenberg-Langevin equation method \cite{Gar85,Car09,Kol07,Ray07,Du23} to investigate the evolution of the photon annihilation operator as a function of propagation length within the medium. This method enables a comprehensive treatment of both coherent and incoherent scattering processes in the atomic gas, the precise ratio of which defines key observables such as second-order correlations.

We begin by examining the atomic polarization for an individual atom interacting with the field, governed by the Langevin equations. These equations are considered within both open and closed system approximations, incorporating the effects of Doppler broadening. We define a closed two-level system as one where the excited state can only decay to the ground state, without additional relaxation processes. In contrast, an open two-level system allows for spontaneous emission to other excited levels, along with collisional relaxation processes, transit-time broadening and other potential decoherence mechanisms. Specifically, the number of atoms interacting with the laser is constant in a closed system, while it fluctuates in an open one. 

The correlation properties of the Langevin forces are described using generalized Einstein relations, providing a robust framework for analyzing the atomic operators. By solving the Langevin equations through successive approximations, involving Fourier transformations and the search for steady-state solutions, we derive an expression for the medium's polarization. This expression is then used to formulate the propagation equation for the photon annihilation operator within the continuous medium approximation. Our analysis includes terms for linear absorption, coherent scattering with biphoton generation, and incoherent scattering arising from the linear and nonlinear components of the Langevin forces.

The final expression for the annihilation operator at the medium's output enables the study of the second-order correlation properties of the transmitted light, as described by the Glauber second-order correlation function. We show that in the specific case of a closed system without Doppler broadening,   the  expressions we derive reproduce known results obtained recently using scattering theory \cite{Mah18,Pra20,Hin21,She23,Kus23,Cor23,Mas23,scatt}, particularly establishing conditions for complete photon anti-bunching.

Beyond validating existing results  our approach allows us to extend  the analysis to significantly more complex systems that are challenging to study using conventional theoretical methods. Specifically, we investigate the influence of Doppler broadening on antibunching conditions in a closed system and examine the consequences of finite light-atom  interaction time for an almost closed system. This comprehensive analysis provides deeper insights into the quantum transport of photons in atomic gases and paves the way for future experimental and theoretical advancements in the field.

Finally, we compare the theoretical predictions with experimental data obtained from a waveguide QED platform, where cold atoms are coupled to the light that is guided through   an optical nanofiber. This platform enables a precise characterization of the photon statistics of the transmitted light under various conditions, including on-resonance and detuned excitation. The theoretical predictions are validated against experimental observations, and   good agreement between theory and experiment is observed.

The paper is structured as follows.
 In Sec. \ref{Sec2},   we derive the Heisenberg-Langevin equations  for the case of  a gas of two-level atoms interacting with a quasi-monochromatic laser field.    The evolution equation for a coordinate-dependent annihilation operator within the Heisenberg representation is obtained. The atomic dynamics are described using Langevin equations, considering the general case of an open two-level system. We also present the correlation properties of the Langevin forces, derived from generalized Einstein relations.
Sec. \ref{Sec3} focuses on solving the Langevin equations for a single atom under the weak saturation approximation. In Sec. \ref{Sec4}, these solutions are employed to derive a propagation equation for the Fourier transform of the annihilation operator, assuming a continuous medium approximation. Sec. \ref{Sec5} provides a detailed investigation of the Glauber two-photon correlation function in various specific scenarios. The comparison between the theoretical model predictions and experimental results is presented in Sec. \ref{Sec6}.
Finally, the main findings and contributions of this study are summarized in the Conclusions.

\section{Heisenberg-Langevin
equation}
\label{Sec2}

\subsection{Model}

 The gaseous atomic   medium consists of two-level atoms that are uniformly distributed with number density \(n\)   within a cylindrical cell with   length \(L\). The gas is not necessarily cold, so Doppler broadening is included. For simplicity, polarization effects are not considered. The medium is assumed to be optically thin in the transverse direction, thereby eliminating any radiation trapping effects \cite{Mol98} in this direction. However, there are no restrictions on the optical depth along the cell's axis.

We consider the evolution of the quantum properties of the input  laser field as it propagates through the given atomic medium. The field is assumed to be nearly monochromatic, with a frequency \(\omega_{p}\)  close to the  resonant frequency  \(\omega_{12}\) of   the  atomic transition.  For simplicity, the   intensity distribution of the field is assumed for all $z$ to be cylindrical, characterized by a uniform intensity across a circular cross-section with radius \( w_0 \)  and with a cross-sectional area \( A = \pi w_0^2 \). Moreover, we neglect  diffraction effects by assuming a small Fresnel number. The influence of diffraction will be discussed in a future publication, where the field will be modeled as a Gaussian beam.

The field is described by quantum mechanical operators in the slowly-varying envelope approximation.
We use the Heisenberg picture, where the operators are time-dependent while the state vector remains time-invariant. In this framework, the dynamics of the system are described by the evolution of the operators. This includes not only the internal interactions within the system but also the coupling of fields to the environment through Langevin force operators. This approach allows us to account for the influence of environmental noise and dissipation on the system's behavior, providing a comprehensive description of the quantum dynamics.

\begin{figure}[tb]
 \centerline{\includegraphics[width=80mm,angle=0]{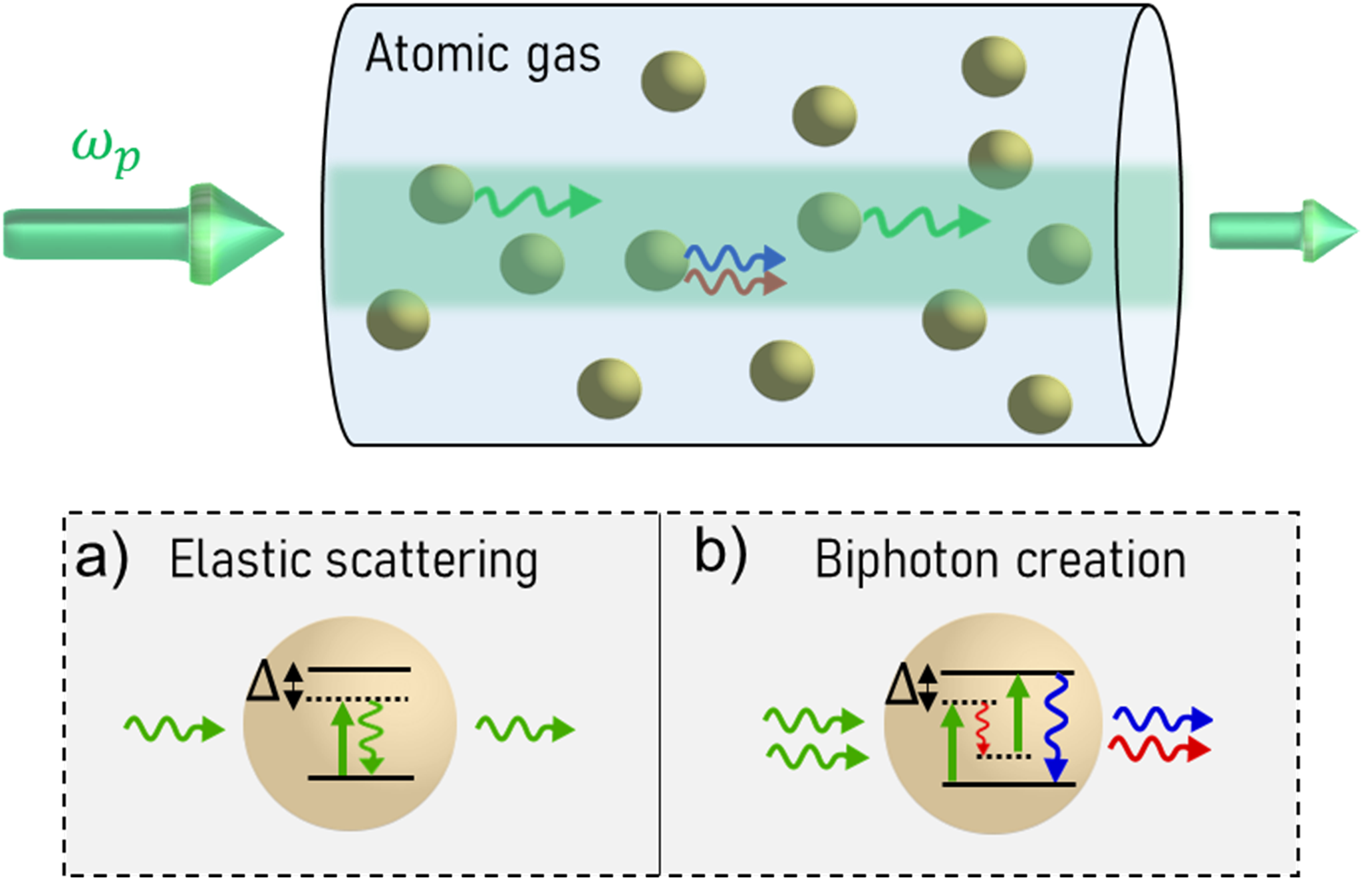}}
 \caption{A nearly monochromatic laser field with frequency \(\omega_{p}\), detuned from the resonant frequency \(\omega_{12}\) of the atomic transition by an amount \(\Delta = \omega_{12} - \omega_{p}\), propagates through an ensemble of two-level atoms.  When the laser is close to resonance,   interaction with the atomic medium can significantly modify the photon statistics at the output. In the    low saturation limit, the two dominant scattering processes are a) elastic scattering and b) biphoton creation. } %\vspace*{20mm}
 \label{fig:Fig1}
 \end{figure}

\subsection{
Continuous-mode quantization}

The electric field  operator $\hat{E}(t ,z)$ of the laser beam as function  of time \(t\) and coordinate $z$ for a radial distance of \(r<w_0\) from the optical axis  can be written as:
\begin{align}
\hat{E}(t ,z)  =& \frac{1}{2}\left[ \hat{E}^{(+)}(t,z) e^{-i\omega_{p}t+ikz}\right.  \nn
& \left.+ \hat{E}^{(-)}(t,z)  e^{i\omega_pt-ikz}\right], \label{field}
\end{align}
where \(\hat{E}^{(\pm)}(t,z)\)  are the amplitudes of the positive and negative frequency components of the electric field as denoted by the \((+)\) and \((-)\) signs in the superscripts. The complex exponentials describe the propagation of the wave  along the positive \(z\) direction with     frequency \(\omega_p\) and    wave vector \(k= \omega_p/c\), where $c$ is the vacuum speed of light.

The Fourier transform  of the amplitude of the positive frequency component of the electric field operator  is given by:
 \be
\hat{E}^{(+)}(\varpi,z) = \frac{1}{\sqrt{2\pi}}\int_{-\infty}^\infty dt \hat{E}^{(+)}(t,z)e^{i\varpi t} .
\ee

We introduce the annihilation operator \(\hat{a}(\varpi,z)\) for a photon with frequency \(\omega_p+\varpi\) at position \(z\) by  the expression
\be
\hat{a}(\varpi,z) =\sqrt{ c\varepsilon_0 A \over 2\hbar \omega_p}\hat{E}^{(+)}(\varpi,z)
 \ee
 relating the positive frequency component of the   electric field, \(\hat{E}^{(+)}(\varpi,z)\) detuned from the carrier frequency \(\omega_p\) by \(\varpi\), to the annihilation operator, \(\hat{a}(\varpi,z)\). The constants \(\hbar\),  \(\varepsilon_0\)    have their usual meanings as the reduced Planck constant  and the vacuum permittivity,  respectively.

In the time domain, the annihilation operator for a photon  at position \(z\) at time \(t\) is determined by the inverse Fourier transform
\begin{equation}
\hat{a}(t,z) =  \frac{1}{\sqrt{2\pi}}   \int d\varpi   \hat{a}(\varpi,z) e^{-i\varpi t}.
\end{equation}
 The annihilation \(\hat{a}(t,z)\) and creation \(\hat{a}^\dag(t,z)\) operators satisfy the canonical commutation relation:
\begin{equation}
[\hat{a}(t,z),\hat{a}^\dag (t',z) ] = \delta(t -t').
\end{equation}

The Fourier transform of the creation operator
\begin{equation}
\hat{a}^\dag(t,z) = \frac{1}{\sqrt{2\pi}}  \int d\varpi   \hat{a}^\dag(\varpi,z) e^{-i\varpi t}
\end{equation}
 is  related to the   operator \(\hat{a}(\varpi,z)^\dag\)      by the identity \(\hat{a}^\dag(\varpi,z)=\hat{a}(-\varpi,z)^\dag\).   
To avoid any confusion, we emphasize that \(\hat{a}^\dag(\varpi,z)\) refers to the Fourier transform of the operator \(\hat{a}^\dag(t,z)\), which is the Hermitian conjugate of \(\hat{a}(t,z)\). In contrast, \(\hat{a}(\varpi,z)^\dag\) denotes the operator that is Hermitian conjugate to \(\hat{a}(\varpi,z)\).

The  operators \(\hat{a}(\varpi,z)^\dag\) and \(\hat{a}(\varpi,z)\) satisfy the  commutation relation:
\begin{equation}
[\hat{a}(\varpi,z) ,\hat{a}(\varpi',z)^\dag ] = \delta(\varpi-\varpi').
\end{equation}

The flow of energy is proportional to the photon flux through the plane at position \(z\) at a given moment \(t\),   measured in terms of photons per unit time. This quantity is described by the flux operator, defined as:
\begin{align}
\hat{f}(t,z) &\equiv \hat{a}^\dagger (t,z)\hat{a}(t,z) \nn
&= \frac{1}{2\pi} \int d\varpi \, d\varpi' \, \hat{a} (\varpi,z)^\dagger\hat{a}(\varpi',z) e^{i( \varpi-\varpi')t}.
\end{align}

It is worth noting that the above description assumes   the pump field has a finite spectral width \(\Delta\varpi_p\), which is much smaller than   all other relevant spectral widths in the problem.
In this context, we assume that the laser beam exhibits a frequency-dependent correlation function, expressed as:
\begin{equation}
\langle \hat{a} (\varpi,z)^\dagger\hat{a}(\varpi',z)\rangle = 2\pi f(\varpi,z) \delta(\varpi-\varpi'),
\end{equation}
where \(f(\varpi,z)\) is the laser power spectrum, defined as the mean photon flux per unit \(\varpi\) bandwidth. We take  the total mean photon flux at a given position \(z\) to remain    constant   over time:
\begin{equation}
  \langle \hat{f}(t,z) \rangle = \int d\varpi f(\varpi,z) \equiv \Phi(z).
\label{flux}
\end{equation}
Here, \(\Phi(z)\) is   time-independent and denotes the mean photon flux at position \(z\), obtained by integrating the laser power spectrum over all frequencies.

\subsection{Propagation equation for annihilation operator}

 Following the approach of Fleischhauer and Lukin \cite{Fle00,Fle02}, we describe the properties of the medium using collective, slowly varying atomic operators \(\hat{\bar{\sigma}}_{ij}(  v_z, t,z)\) (\(i, j = 1, 2\)), which are defined as:
\be
\hat{\bar{\sigma}}_{ii}( v_z, t,z) = \frac{1}{N(v_z, z)} \sum_{\nu=1}^{N(v_z, z)} \hat{\sigma}^{(\nu)}_{ii}, \label{operii}
\ee
\be
\hat{\bar{\sigma}}_{12}( v_z, t,z) = \frac{1}{N(v_z, z)} \sum_{\nu=1}^{N(v_z, z)} \hat{\sigma}^{(\nu)}_{12} e^{+i(\omega_p t - kv_z)}, \label{oper12}
\ee
\be
\hat{\bar{\sigma}}_{21}( v_z, t,z) = \frac{1}{N(v_z, z)} \sum_{\nu=1}^{N(v_z, z)} \hat{\sigma}^{(\nu)}_{21} e^{-i(\omega_p t- kv_z)}, \label{oper21}
\ee
where \( \hat{\sigma}^{(\nu)}_{ij} = |i_\nu\rangle \langle j_\nu| \) is the atomic operator for a single atom \(\nu\), and \(\omega_p - kv_z\) is the laser frequency in the frame of reference moving with the atom, with \( kv_z \) representing the Doppler shift. The averaging in Eqs.~(\ref{operii})--(\ref{oper21}) is performed over thin but macroscopic spatial layers of width \(dz\), denoted by their position \(z\), and over a narrow, but finite, velocity interval \(dv_z\) near velocity \(v_z\), containing a large number of atoms, 
\[
N (v_z,z) = n A W(v_z) dz dv_z \gg 1,
\]
where
\be
W(v_z) = \frac{1}{\sqrt{\pi} v_0} \exp\left(-\frac{v_z^2}{v_0^2}\right)
\ee
is the Maxwell velocity distribution, and \(v_0 = \sqrt{\frac{2k_B T}{m}}\) is the most probable velocity. Here, \(k_B\) is the Boltzmann constant, \(T\) is the temperature, and \(m\) is the atomic mass.

The evolution of the annihilation operator with respect to position \(z\) in the atomic medium is described by the  propagation equation \cite{Du23}
\begin{equation}
\frac{\partial \hat{a}(\varpi,z)}{\partial z} = i g \int d v_z n A W(v_z) \hat{\bar{\sigma}}_{1  2   }(v_z,\varpi,z),
\label{prop0}
\end{equation}
where \(\hat{\bar{\sigma}}_{1 2   }(v_z,\varpi,z)\) is the Fourier transform of the collective atomic coherence operator  \(\hat{\bar{\sigma}}_{1  2   }(v_z, t,z)\).
The coupling strength \(g \) is given by
\be
g =\mu  \sqrt{\omega_{12}/(2 c \varepsilon_0 \hbar A)},\ee
where \(\mu \) is the dipole moment of the atomic transition, which we assume to be real and positive.

 The expression for the coupling strength \(g\) can be written in a simpler form, which is clearer from a physical point of view:
\be
g = \sqrt{\beta \Gamma},
\ee
where \(\Gamma\) is the rate of spontaneous transitions from the upper state \(|2\rangle\) to the lower state \(|1\rangle\), and \(\beta\) is the fraction of photons emitted into the target mode. For free-space propagation, one gets \cite{Gro82}:
\be
\beta = \frac{3\lambda^2}{8\pi^2 w_0^2} \ll 1.
\ee
Here, \(\lambda = 2\pi c / \omega_{12}\) is the transition wavelength.

\subsection{Single two-level atom Langevin equations}

 In a reference frame rotating with the Doppler-shifted laser frequency $\omega_p -k v_z$, the interaction Hamiltonian for a single atom moving with velocity \(v_z\) is given by
\begin{equation}
 \hat{V} = -\hbar g\left( \hat{\sigma}_{21}\hat{a} +   \hat{a}^\dag \hat{\sigma}_{12}\right) + \hbar(\Delta + k v_z)\hat{\sigma}_{22}, \label{Ham}
  \end{equation}
 where \(\Delta = \omega_{12} - \omega_p\) is the detuning from resonance for an unmoving atom.

The atomic evolution is governed by the
 Langevin equation
\begin{align}
{\partial \over \partial t}\hat{\sigma}_{jk}={i\over \hbar} [\hat{V},\hat{\sigma}_{jk}]+\hat{L}_{jk}+ \hat{f}_{jk} ,
\label{eq1}
\end{align}
where  \(\hat{L}_{jk}\) describe the relaxation and pumping processes (spontaneous emission, atomic collisions, transit-time broadening and so on),
and  \(\hat{f}_{jk} \) are   the
  Langevin noise operators \cite{Kol07,Du23}.
 
\begin{figure}[t]
	\centerline{\includegraphics[width=85mm,angle=0]{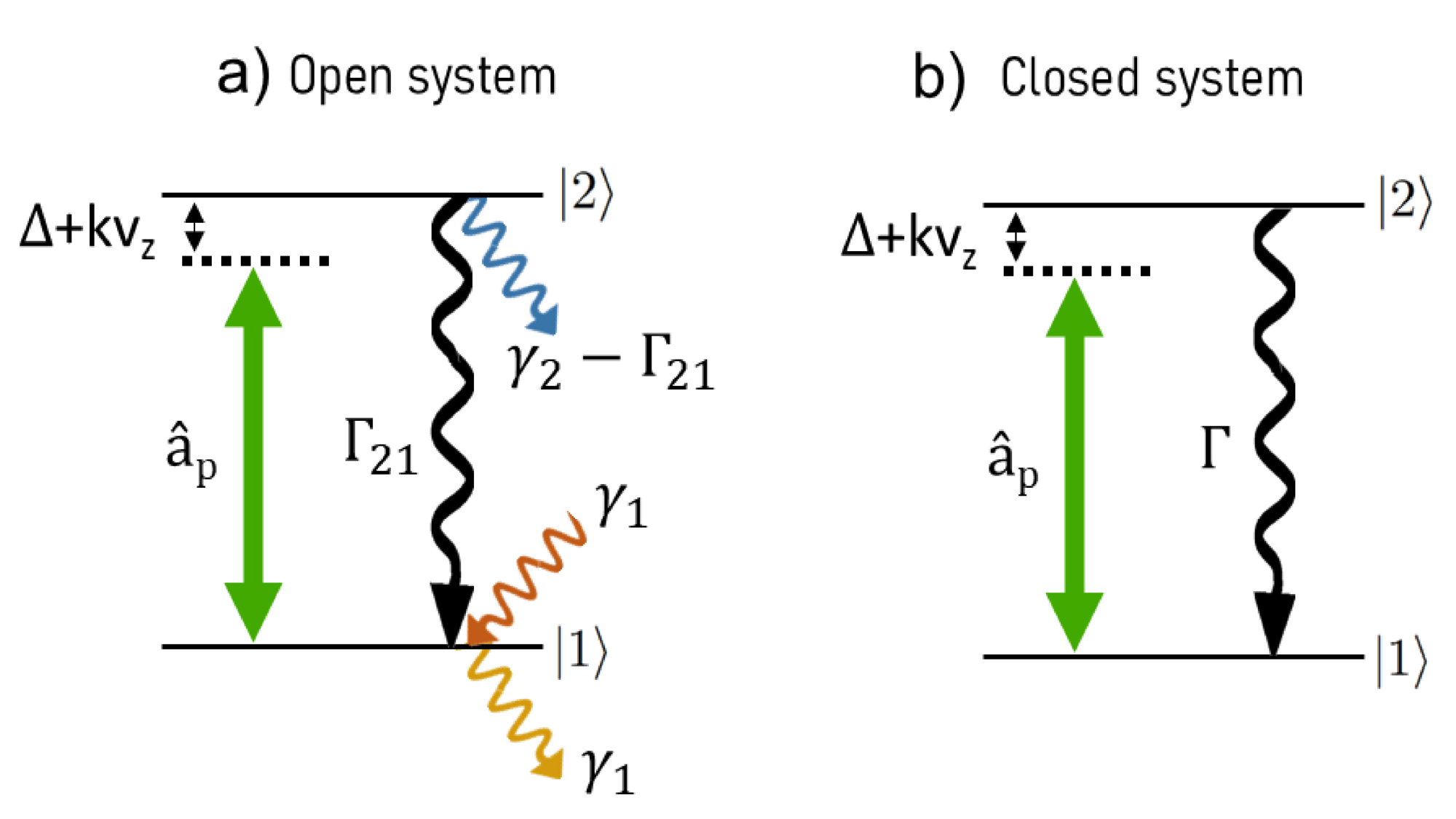}}
	\caption{Model of relaxation processes in an open two-level system (a) and in a closed system (b). In the open system, state \(|1\rangle\) decays out of the system at a rate \(\gamma_{1}\) and is incoherently pumped at the same rate; state \(|2\rangle\) decays with a total rate \(\gamma_{2}\)  consisting of    the rate \(\Gamma_{21}\) of population transfer from state \(|2\rangle\) to level 1 due to spontaneous transitions, inelastic collisions, and other processes  and decay out of the system at a rate \(\gamma_{2} - \Gamma_{21}\). The coherence between the states decays at a rate \(\gamma_{12} \geq (\gamma_{1} + \gamma_{2}) / 2\), which is not depicted in the figure. In the case of the closed system, the upper state decays only spontaneously to the lower state at a rate \(\Gamma\), and the decoherence rate is \(\gamma_{12} = \Gamma / 2\).}  
	%\vspace*{20mm}
	\label{fig:levels}
\end{figure}

Using the Hamiltonian (\ref{Ham}) and assuming the simplest model to describe relaxation processes in an open two-level system with relaxation constants (see Fig. \ref{fig:levels}), the Eqs.~(\ref{eq1}) for the atomic operators of a single atom can be explicitly written as follows:
\begin{widetext}
\bea
{\partial \over \partial t}\hat{\sigma}_{1 2 }(t)&=& - ig  \hat{a}(t)\left[\hat{\sigma} _{2 2 }(t) -\hat{\sigma}_{1 1}(t) \right]
 -\left(\gamma_{12}  -i\Delta-i k v_z\right)\hat{\sigma}_{1 2 }(t)+\hat{f}_{1 2 }(t)  ,   \label{eq12}\\
 {\partial \over \partial t}\hat{\sigma}_{21 }(t)&=&  ig \left[\hat{\sigma} _{2 2 }(t) -\hat{\sigma}_{1 1}(t) \right]\hat{a}^\dag(t)\
 -\left(\gamma_{12}  +i\Delta+i k v_z\right)\hat{\sigma}_{21 }(t)+\hat{f}_{21 }(t)  ,\label{eq21}\\
 {\partial \over \partial t}\hat{\sigma}_{1 1 }(t)&=&-i g \left[\hat{\sigma}_{2 1 }(t)\hat{a}(t) -    \hat{a}^\dag(t) \hat{\sigma}_{1 2 }(t) \right]
 +\Gamma_{21} \hat{\sigma}_{2 2 }(t) - \gamma_1 \left(\hat{\sigma}_{1 1 }(t)-  1\right)+\hat{f}_{1 1 }(t) , \label{eq11}\\
 {\partial \over \partial t}\hat{\sigma}_{22 }(t)&=&i g\left[ \hat{\sigma}_{2 1 }(t) \hat{a}(t)-    \hat{a}^\dag(t) \hat{\sigma}_{1 2 }(t) \right]
   - \gamma_2  \hat{\sigma}_{22 }(t) +\hat{f}_{22 }(t) \label{eq22}.
\eea
\end{widetext}
Here, \(\gamma_{12}\) is the coherence relaxation rate, \(\gamma_i\) is the population relaxation rate of level \(i\), and \(\Gamma_{21}\) is the rate of population transfer from level 2 to level 1 due to spontaneous transitions, inelastic collisions, and other processes, with \(\Gamma_{21} < \gamma_2\). In the following, we will assume that the primary contribution to this transfer comes from spontaneous emission, resulting in \(\Gamma_{21} = \Gamma\).

 The special case of the closed system, which will be discussed at several points throughout the article, can be obtained from the results derived using Eqs.~(\ref{eq12})--(\ref{eq22}) for the open system by substituting
\be
\gamma_1 = \gamma, \quad \gamma_2 = \Gamma + \gamma, \quad \gamma_{12} = \frac{\Gamma}{2} + \gamma, \label{closed}
\ee
and then taking the limit \(\gamma \rightarrow 0\). 

In Fourier space, using  the fact that  the Fourier transform of a product of operators \( \hat{P}(t) = \hat{A}(t) \hat{B}(t)\) can be expressed as
\be
  P(\varpi)={1\over \sqrt{2\pi}}\int d\nu\hat{A}(\nu)\hat{B}(\varpi-\nu),
\ee
the Langevin equations   (\ref{eq12})--(\ref{eq22}) for steady-state conditions  read
\begin{widetext}
\bea
0&=& -i    {g\over \sqrt{2\pi}}\int d\nu \hat{a}(\nu)\left[\hat{\sigma} _{22}(\varpi-\nu) -\hat{\sigma}_{11}(\varpi-\nu)\right]
 -G_{12}(\varpi)\hat{\sigma}_{12}(\varpi)+\hat{f}_{12} (\varpi), \label{eq24}\\
%%%%%%%%%%%
0&=& +i    {g\over \sqrt{2\pi}}\int d\nu \left[\hat{\sigma} _{22}(\varpi-\nu) -\hat{\sigma}_{11}(\varpi-\nu)\right]\hat{a}^\dag(\nu)
 -G_{21}(\varpi)\hat{\sigma}_{21}(\varpi)+\hat{f}_{21}(\varpi),\\
%%%%%%%%%%%
0&=&-i    {g\over \sqrt{2\pi}}\int d\nu \left[\hat{\sigma} _{21}(\varpi-\nu)\hat{a}(\nu) - \hat{\sigma} _{12}(\varpi-\nu)\hat{a}^\dag(\nu)\right] + \Gamma \hat{\sigma}_{22}(\varpi) -   G_{1}(\varpi)[\hat{\sigma}_{11}(\varpi) -\sqrt{2\pi}   \delta(\varpi)]+\hat{f}_{11} (\varpi),\\
%%%%%%%%%%%%%%%
0&=&+i     {g\over \sqrt{2\pi}}\int d\nu\left[\hat{\sigma} _{21}(\varpi-\nu)\hat{a}(\nu)  -\hat{a}^\dag(\nu) \hat{\sigma} _{12}(\varpi-\nu) \right]   -G_{2}(\varpi)\hat{\sigma}_{22}(\varpi)+\hat{f}_{2 2} (\varpi), \label{eq26}
\eea
\end{widetext}
where we introduce complex rates \begin{align}
 G_{12}(\varpi)&=\gamma_{12} - i\Delta - ikv_z - i\varpi ,\label{G12}\\
 G_{21}(\varpi)&=\gamma_{12} + i\Delta +ikv_z - i\varpi ,\\  G_{j}(\varpi)&=\gamma_j - i\varpi .
\end{align}

\subsection{Langevin forces }

The operators \( \hat{f}_{\alpha}(t) \), describing the Langevin noise forces, and their Fourier transforms \( \hat{f}_{\alpha} (\varpi) \) are \(\delta\)-correlated:
\bea
\langle\hat{f}_{\alpha} (t) \hat{f}_{\alpha'} (t')\rangle_a&=&\hat{\mathcal{D}}_{\alpha,\alpha'}\delta(t-t'),\\
\langle\hat{f}_{\alpha} (\varpi  ) \hat{f}_{\alpha'} (\varpi')\rangle_a&=&\hat{\mathcal{D}}_{\alpha,\alpha'}\delta(\varpi+\varpi'),
\eea
where the indices \(\alpha\) take the values
\be \alpha=\{1,2\},  \{2,1\}, \{1,1\}, \{2,2\} ,\label{order}\ee 
and $\langle\,\,\,\rangle_a$ denotes the averaging over atomic variables. 
As described in \cite{Du23}, the diffusion coefficients \(\hat{\mathcal{D}}_{\alpha,\alpha'} \) can be obtained using generalized Einstein relations
\be
\hat{\mathcal{D}}_{\alpha,\alpha'}={\partial \over \partial t}\langle \hat{\sigma}_{\alpha}\hat{\sigma}_{\alpha'}\rangle_a-\langle \hat{A}_{\alpha}\hat{\sigma}_{\alpha'}\rangle_a
-\langle \hat{\sigma}_{\alpha}\hat{A}_{\alpha'}\rangle_a ,
\ee
where \(\hat{A}_{\alpha} ={\partial \over \partial t}\hat{\sigma}_{\alpha}-\hat{f}_{\alpha}\).
 For an open two-level system, the diffusion coefficients can be expressed in the following matrix form:
 \begin{widetext}
\bea  \label{Diff-matrix}
&&\hat{\mathcal{D}}  =   \begin{bmatrix}
0 &2    \gamma_{12}\hat{s}_{11}    +\Gamma\hat{s}_{22}+\gamma_1(1-\hat{s}_{11}) &  -(\gamma_1+\Gamma )  \hat{s}_{12} &  \gamma_2  \hat{s}_{12}   \\
 ( 2\gamma_{12} -\gamma_{ 2} )\hat{s}_{22}  &0&  0 &  0   \\
0& -(\gamma_1+\Gamma)\hat{s}_{21}   &  \Gamma\hat{s}_{22}+\gamma_1(1-\hat{s}_{11}) &  -(\gamma_1+\Gamma) \hat{s}_{22}    \\
0& \gamma_2  \hat{s}_{21}   &-(\gamma_1+\Gamma )\hat{s}_{22}    &     \gamma_2\hat{s}_{22}  \\
\end{bmatrix}.
\eea
\end{widetext}
Here, we have followed the ordering given in (\ref{order})   for the indices 
$\alpha$ and $\alpha'$.
The operators \( \hat{s}_{ij}=  \langle \hat{\sigma}^{(0)}_{ij}\rangle_a\)  are  the  stationary solutions of Eqs. (\ref{eq12})--(\ref{eq22}) for \(\hat{f}_{\alpha}=0\) and a time-independent field annihilation operator $\hat{a}$. These operators, derived under the weak saturation approximation \(S(z) \ll 1\), 
are provided in Appendix \ref{appendix1} (see Eqs.~(\ref{sigma01})--(\ref{sigma02})).
We define the saturation parameter \(S(z)\) in terms of the photon flux (\ref{flux}) and relaxation parameters for the general case of an open system as:
\begin{equation}
	S(z) = 2\beta \frac{\Gamma (\gamma_1 + \gamma_2 - \Gamma)}{\gamma_{12} \gamma_1 \gamma_2} \Phi(z).
	\label{sat_par}
\end{equation}
For a closed system, it takes the form:
\begin{equation}
	S(z) = 8\beta \frac{\Phi(z)}{\Gamma}.
	\label{sat_par_closed}
\end{equation}

\section{Solution of Langevin atomic equation for weak saturation}

\label{Sec3}

We are focused on the quantum effects occurring during the propagation of   nearly monochromatic laser radiation (with a linewidth \(\Delta \varpi_p\) much smaller than all characteristic spectral parameters of the problem) in a resonantly absorbing gas of two-level atoms.   Within the narrow detuning range \(|\varpi| \leq \Delta \varpi_p\), the annihilation operator \(\hat{a}(\varpi, z)\) characterizes the properties of the nearly coherent pump laser radiation propagated to position $z$. We will denote this by the operator:
\be
\hat{a}_{p}(z) = {1\over \sqrt{2\pi}}\int_{\Delta \varpi_p} d\varpi \, \hat{a}(\varpi, z).
\label{aP}
\ee

For detunings \(|\varpi| > \Delta \varpi_p\), the annihilation operator at the entrance to the medium is
\be \hat{a}(\varpi, 0) = \hat{b}(\varpi) . \label{vac_b}\ee
 The field annihilation operators \(\hat{b}(\varpi)\) have the commutation relation \([\hat{b}(\varpi), \hat{b}(\varpi')^\dag] = \delta(\varpi - \varpi')\), and since they act only on the vacuum state, the mean photon flux at the entrance to the medium for frequencies \(|\varpi| > \Delta \varpi_p\) is zero: \(\langle \hat{b}(\varpi)^\dag \hat{b}(\varpi) \rangle = 0\).

The pump field described in the medium \(0<z<L\) by the operator 
\(
\hat{a}_{p}(z)\)
 is assumed to be weak and far below saturation: $S(z)
\ll 1$. In this regime, it is common to decompose the medium response into   linear and   nonlinear contributions. 

Let us first consider the medium's linear response. The primary effect here is linear absorption. In the language of photon scattering, linear absorption is described by the   elastic forward scattering of the photon   into the spatial mode of the pump field (see Fig.~\ref{fig:Fig1}a), with the field attenuation caused by a  phase shift (\(\pi\) at exact resonance), acquired by the   scattered photon \cite{Tan11}.

 In the Heisenberg picture, the pump annihilation operator \(\hat{a}_{p}(z)\) decreases exponentially with \(z\) due to linear absorption. This might appear to violate the commutation relations if not accounted for properly. However, the quantum description inherently ensures that these relations are preserved. This is achieved by suitable adjustments of the annihilation operator \(\hat{a}(\varpi, z)\) for detunings \(\varpi\) much larger than the laser linewidth. These modifications reflect the interaction of the vacuum field with fluctuations in the medium's linear polarization, represented by the Langevin force \(\hat{f}_{12}\). Importantly, no real photons are generated at these frequencies, and the radiation spectrum remains unchanged as the nearly monochromatic light propagates through the linearly absorbing medium.

When one considers nonlinear effects in the interaction between the pump field and two-level atoms, more complex processes can occur, one of which is illustrated in Fig.~\ref{fig:Fig1}b. In this process, two photons from the pump field are absorbed, and two photons with detunings \(\varpi_1 = -\varpi_2\) are emitted back into the spatial mode of the pump field \cite{Dal83, Mah18, Pra20, Hin21, She23, Kus23, Cor23, Mas23, scatt}. As a result, the spectrum of the light broadens, and its coherence properties, described by higher-order correlation functions, change significantly.

Overall, to describe these linear and nonlinear processes within the outlined model, for small frequencies \(|\varpi| < \Delta \varpi_p\), it is sufficient to know the atomic polarization operator \(\hat{\sigma}_{12}(\varpi)\) in the linear approximation with respect to \(\hat{a}_{p}\). For large detunings \(|\varpi| > \Delta \varpi_p\), the description requires a linear approximation with respect to \(\hat{a}(\varpi)^\dag\) and the Langevin forces \(\hat{f}_{j}(\varpi)\), and a second-order approximation with respect to \(\hat{a}_{p}\).

The solution of the Langevin equations for a single atom can be generally expressed as
\be \hat{\sigma}_{\alpha}(t) = \hat{\sigma}^{d}_{\alpha}(t) + \sum_{\alpha'} \hat{A}_{\alpha,\alpha'} \hat{f}_{\alpha'} (t).
\label{sum}
\ee
 The dynamical term \(\hat{\sigma}^{d}_{\alpha}(t)\) is the solution of the Eqs.~(\ref{eq24})--(\ref{eq26})
    for \(\hat{f}_{\alpha}=0\). It is the same for all atoms in the small averaging volume  \( dz dv_z\) and describes the coherent contribution to the scattering of the pump field along the \(z\) coordinate. The incoherent term \(\sum_{\alpha'} \hat{A}_{\alpha,\alpha'} \hat{f}_{\alpha'}\) describes the contribution of the Langevin fluctuations.

Under the aforementioned assumptions, the steady-state solution for the atomic coherence operator of a single atom in the weak saturation approximation \(S(z) \ll 1\) is given by a sum in Eq. (\ref{sum}), with the dynamical part

\begin{align}
\hat{\sigma}^d_{12}(\varpi) &= i \frac{g}{G_{12}(\varpi)} \hat{a}(\varpi) \nn \\ \nonumber
&\quad - i \frac{g^3 \hat{a}_{p} \hat{a}_{p}}{G_0(\varpi) G_{12}(\varpi)} \left[ \frac{1}{G_{21}(\varpi)} + \frac{1}{G_{12}(0)} \right] \hat{a}(-\varpi)^\dag
\end{align}
and the   noise contribution \(\sum_{\alpha'} \hat{A}_{12,\alpha'}(\varpi) \hat{f}_{\alpha'} (\varpi)\) with coefficients
\begin{align}
\hat{A}_{12,12}(\varpi) &= \frac{1}{G_{12}(\varpi)}, \\
\hat{A}_{12,21}(\varpi) &= \frac{g^2 \hat{a}_{p} \hat{a}_{p}}{G_{12}(\varpi) G_{21}(\varpi) G_0(\varpi)}, \\
\hat{A}_{12,11}(\varpi) &= \frac{i g \hat{a}_{p}}{G_{12}(\varpi) G_1(\varpi)}, \\
\hat{A}_{12,22}(\varpi) &= - \frac{i g \hat{a}_{p}}{G_{12}(\varpi) G_2(\varpi)} \left[ 1 - \frac{\Gamma}{G_{11}(\varpi)} \right],
\end{align}
where \(G_0(\varpi) = \left[ \frac{1}{G_1(\varpi)} + \frac{1}{G_2(\varpi)} - \frac{\Gamma}{G_1(\varpi) G_2(\varpi)} \right]^{-1}\).

In deriving the dynamic part, we retained only the terms describing the generation of biphotons and omitted the irrelevant terms that describe the slight change in absorption due to the saturation effect.

\section{ Propagation equation and its solution}
\label{Sec4}
\subsection{Derivation of the propagation equation}

The obtained expressions for the polarization of a single atom enable us to derive the expression for the collective operator \(\hat{\bar{\sigma}}_{1  2   }\), which defines the propagation equation (\ref{prop0}).
Since the Langevin noise contributions  of different atoms in the averaging volume are completely random and uncorrelated, the collective operators \(\hat{\bar{\sigma}}_{\alpha}\) can be written as \cite{Du23}:
\begin{align}
\hat{\bar{\sigma}}_{\alpha}(t, z,v_z) =& \hat{\sigma}^d_{\alpha}(t, z,v_z) +\nn
&  \frac{1}{\sqrt{n A W(v_z)}} \sum_{\alpha'} \hat{A}_{\alpha,\alpha'} \hat{\bar{f}}_{\alpha'}(t, z,v_z),
\end{align}
where \(\hat{\bar{f}}_{\alpha'}(t, z,v_z)\) is delta-correlated not only in time \(t\) but also in the coordinate \(z\) and in velocity \(v_z\):
\begin{align}
\langle\hat{\bar{f}}_{\alpha} (t, z) \hat{\bar{f}}_{\alpha'} (t', z')\rangle_a &= \hat{\mathcal{D}}_{\alpha,\alpha'} \delta(t - t') \delta(z - z')\delta(v_z - v_z'),
\end{align}
and the diffusion coefficients \(\hat{\mathcal{D}}_{\alpha,\alpha'}\) are given by Eq.~(\ref{Diff-matrix}).

We will divide the frequency space into the very narrow interval near zero given by the probe laser linewidth   \(\Delta \varpi_p\) and an interval containing all other frequencies. In the narrow interval, we will neglect the contribution of biphoton generation and consider only the linear absorption of the pump field. In the second interval, the annihilation operator can be expressed in terms of contributions from the modified vacuum part due to linear Langevin noise, as well as from biphoton generation due to both the nonlinear dynamical part and the nonlinear Langevin contribution.

In the narrow frequency interval, the propagation equation (\ref{prop0}), combined with the linear atomic coherence \(\hat{\sigma}^{(lin)}_{12}(\varpi) = i g \hat{a}(\varpi)/G_{12}(\varpi)\), yields the propagation equation for the pump field annihilation operator \(\hat{a}_{p}(z)\):  
\be  
\frac{\partial \hat{a}_{p}(z)}{\partial z} = -\frac{1}{2}\alpha(0) \hat{a}_{p}(z),  
\label{prop_aP0}  
\ee  
where the complex absorption coefficient \(\alpha(\varpi)\) for a weak field detuned by \(\Delta + \varpi\) from the center of the Doppler-broadened line is defined using \(G_{12}(\varpi)\) [Eq.~(\ref{G12})]:  
\begin{equation}  
	\alpha(\varpi) = \alpha_0 \int dv_z W(v_z) \frac{\gamma_{12}}{\gamma_{12} - i(\Delta + \varpi) - i k v_z}.  
	\label{absorption}  
\end{equation}  
Here, \(\alpha_0\) is the linear absorption at the line center in the absence of Doppler broadening, where \(W(v_z) = \delta(v_z)\).

In the wide frequency range, we have:
\begin{align}
{\partial \hat{a}(\varpi,z)\over \partial z}=&-{1\over 2}\alpha(\varpi)\hat{a}(\varpi,z) + \hat{\chi}(\varpi,z)   \hat{a}(-\varpi,z)^\dag   \nn&+\hat{F}_{l}(\varpi,z) +   \hat{F}_{nl}(-\varpi,z)^\dag ,
\label{prop_NL}
\end{align}
where
%\begin{widetext}
the nonlinear contribution to the dynamical part of polarization is described by
\begin{align}
\hat{\chi}(\varpi,z)=& \beta\alpha_0 \int d v_z W(v_z)\times \nn & {\Gamma\gamma_{12}[G_{21}(\varpi)+G_{12}(0)] \hat{a}_{p}(z)\hat{a}_{p}(z)\over
 2G_0(\varpi) G_{12}(0)G_{12}(\varpi)G_{21}(\varpi) } .
 \label{eq_wide0}
\end{align}
%\end{widetext}
From Eq.~\eqref{eq_wide0}, one can see that term $ \hat{\chi}(\varpi,z)   \hat{a}(-\varpi,z)^\dag $ in Eq.~\eqref{prop_NL} describes the annihilation of two pump photons and the creation of a photon at frequency $-\varpi$. Finally, this term will describe biphoton generation as depicted in Fig.~\ref{fig:Fig1}b.

The linear and nonlinear  parts of the Langevin noise contribution in (\ref{prop_NL}) are defined as
 \begin{align}
\hat{F}_{l}&(\varpi,z) = i\int dv_z \sqrt{{\alpha_0 W(v_z)\gamma_{12}\over 2} } \frac{\hat{f}_{12}(\varpi,z,v_z) }{G_{12}(\varpi)},\\
  \hat{F}_{nl}&(-\varpi,z)^\dag= i\int dv_z \sqrt{{\alpha_0 W(v_z)\gamma_{12}\over 2} }\times\nn &
 \left[ \hat{A}_{12,21}(\varpi)\hat{f}_{21}(\varpi,z,v_z) + \hat{A}_{12,11}(\varpi)\hat{f}_{11}(\varpi,z,v_z) \right. \nn&+\left. \hat{A}_{12,22}(\varpi)\hat{f}_{22}(\varpi,z,v_z)  \right] .
\label{Fnl}
\end{align}

Since the contribution of the nonlinear part to absorption is negligible, we set \(\hat{F}_{nl}(\varpi)  |0\rangle =0\), where \(|0\rangle\) is the initial state of the system with only the pump light present. 
Thus, the nonlinear part of Langevin contribution describes only the annihilation of two pump photons and the    creation of a photon with frequency \( - \varpi\), \(\hat{F}_{nl}( \varpi)^\dag|0\rangle\neq 0\), by atoms located near a point \(z\) and having a velocity near \(v_z\). 

For the operator \( \hat{F}_{l}(\varpi) \),   we have not only  \(\hat{F}_{l}(\varpi)^\dag|0\rangle\neq 0\)  but also \(\hat{F}_{l}(\varpi) |0\rangle\neq 0\) due to existing correlation between \(\hat{f}_{21}\) and  \(\hat{f}_{12}\), specifically  (\(\hat{\mathcal{D}}_{21,12}=(2\gamma_{12}-\gamma_2)  \hat{ s }_{22} \neq 0 )\).

The creation of real photons due to Langevin fluctuations of the coherence operator \(\hat{\sigma}_{12}\) takes place    if two conditions are satisfied: there is some saturation (\(  \hat{ s} _{22}  \neq 0\)), and the system is not   an ideal closed system (\(\gamma_{12} > \gamma_2 / 2\)). 

In scenarios involving an open system or a closed system with additional decoherence mechanisms (such as phase-changing collisions in atomic gases or phonon decoherence in rare earth ions doped in crystals),  the correlation term \(\hat{\mathcal{D}}_{21,12}\) describes the process whereby, due to the absorption of pump photons, a nonzero population of the excited state is generated. Decoherence  will cause the phases of the excited states of different atoms to be (partially) uncorrelated leading to the independent emission  of spontaneous photons. This creates   a thermal-like radiation with a spectral width of  \(\gamma_{12}\). A fraction \(\beta\) of these photons is emitted in the spatial mode of the pump field. We will refer to these photons as thermal-like spontaneous  photons.

\subsection{Perturbation solution of propagation  equation }

The solution of Eq. (\ref{prop_aP0}) is straightforward:
\be
\hat{a}_{p}(z)=e^{-\frac{1}{2}\alpha(0) z}\hat{a}_{p}(0),
\ee
leading to the following expression for the annihilation operator   near \(\varpi=0\):
\be
\hat{a} (\varpi,z)  \approx\sqrt{2\pi}\hat{a}_{p}(z)\delta(\varpi).
\label{solution_pump}
\ee

 Equation (\ref{prop_NL}) can be solved using the method of successive approximations, treating the term \(\hat{\chi}(\varpi, z) \hat{a}(-\varpi, z)^\dag + \hat{F}_{nl}(-\varpi, z)^\dag\) as a small perturbation.

The zeroth-order solutions for the annihilation operator \(\hat{a} ^{(0)}(\varpi,z)\) in the wide interval \(|\varpi| > \Delta \varpi_p\) satisfy the equation
\begin{align}
\frac{\partial \hat{a} ^{(0)}(\varpi,z)}{\partial z} &= -\frac{1}{2}\alpha(\varpi) \hat{a}^{(0)}(\varpi,z) + \hat{F}_{l}(\varpi,z) \label{eq_zero}
\end{align}
with initial conditions
\be
\hat{a} ^{(0)}(\varpi,0)=\hat{b}(\varpi).
\ee

The solution of this equation is
\begin{align}
\hat{a} ^{(0)}(\varpi,z) &\equiv B (\varpi,z)= e^{-\frac{1}{2}\alpha(\varpi) z}\hat{b}(\varpi) \nn
&+ \int_0^z d\zeta e^{-\frac{1}{2}\alpha(\varpi)(z-\zeta)}\hat{F}_{l}(\varpi,\zeta).
\label{zero_order}
\end{align}

The first-order term \(\hat{a} ^{(1)}(\varpi,z)\) satisfies the equation
\begin{align}
{\partial \hat{a}^{(1)}(\varpi,z)\over \partial z}=&-{1\over 2}\alpha(\varpi)\hat{a}^{(1)}(\varpi,z)  \nn &+     \hat{\chi}(\varpi )   \hat{B}(-\varpi,z)^\dag  + \hat{F}_{nl}(-\varpi,z)^\dag 
\label{prop_first}
\end{align}
with the initial condition
\be
\hat{a} ^{(1)}(\varpi,0)= 0.
\ee
The solution of this equation is
%\begin{widetext}
\begin{align}
\hat{a} ^{(1)}(\varpi,z)&\equiv \hat{C}(-\varpi,z)^\dag= \int_0^z d\zeta  e^{- \frac{1}{2}\alpha(\varpi)(z-\zeta)} \nn&  \times \left[ \hat{\chi}(\varpi,\zeta ) \hat{B}(-\varpi,\zeta)^\dag
+    \hat{F}_{nl}(-\varpi,\zeta)^\dag\right].
\label{first_order}
\end{align}
%\end{widetext}

Thus,  we obtain the following solution for the annihilation operator:
\begin{align}
\hat{a}(\varpi,z) &=\sqrt{2\pi}\hat{a}_{p} ( z)\delta(\varpi) +\hat{B} (\varpi,z) +  \hat{C} (-\varpi,z)^\dag.
\label{solution}
\end{align}
Recall that $\hat{B}$ describes the linear response (absorption and linear Langevin fluctuations) and $\hat{C}$ describes the nonlinear response (coherent biphotons generation and nonlinear Langevin fluctuations).

 \section{Two-photon intensity correlation function }
 \label{Sec5}
 \subsection{General expressions}
 \label{subsec_Gen_exp}

 The  expression (\ref{solution}) with (\ref{zero_order}) and (\ref{first_order}) for the annihilation operator enables us to calculate characteristics of the field after its  interaction with a gas of two-level atoms. In this paper, we will limit ourselves to examining the time correlation properties of the transmitted field. Thus, we will analyze the Glauber two-photon correlation function of the output field, defined as
\be
G^{(2)}(\tau) = \langle \hat{a}^\dag(t,L)\hat{a}^\dag(t+\tau,L) \hat{a}(t+\tau,L)\hat{a}(t,L)\rangle.
\label{G2_gen}
\ee

The solution (\ref{solution}) for the annihilation operator is valid in the first-order approximation with respect to the parameter \(\beta\). For the expression for the correlation function \(G^{(2)}(\tau)\) to be accurate up to the order of \(\beta^2\), we generally need the solution for the annihilation operator in the second-order approximation with respect to \(\beta\). However, in this work, for simplicity, we will consider only a sufficiently weak field entering the medium. In this case, the main contribution to \(G^{(2)}(\tau)\) comes from the term proportional to \(\Phi_0^2\), where  \(\Phi_0 = \langle \hat{a}_{p}(0)^\dag \hat{a}_{p}(0) \rangle\) is the pump photon flux at the entrance of the medium. This term is much larger than the terms proportional to \(\Phi_0^3\) and higher. It is easy to show that in this case the solution (\ref{solution}) is sufficient to determine  \(G^{(2)}(\tau)\) with an accuracy of the order of \(\beta^2\).   A more detailed analysis of the correlation properties in higher orders of \(\beta\) will be carried out in a subsequent publication.

Details of the derivation of the expressions for the second-order intensity correlation function \( G^{(2)}(\tau) \) are presented in the appendix \ref{appendix}.
In the weak field approximation, the second-order intensity correlation function \( G^{(2)}(\tau) \) consists of two parts: 
\begin{equation}
G^{(2)}(\tau) = G^{(2)}_b(\tau) + G^{(2)}_s(\tau).
\label{G2general}
\end{equation} 

The first term describes the contribution of      pump   field attenuated by linear absorption and     biphotons field to $G^{(2)}(\tau)$:
\begin{align}
G^{(2)}_b(\tau)
=& \Phi_0^2  \left| e^{-\alpha(0) L} + \beta \psi_b(\tau) \right|^2,
\label{G2biph}
\end{align}
where   $\psi_b(\tau)$ is defined as the biphoton wavefunction.
Note that despite well-defined mutual second order coherence, the transmitted pump and biphoton fields do not exhibit  mutual first-order coherence.

The second term of Eq.~\eqref{G2general} describes the contribution of the thermal-like spontaneous photons   to $G^{(2)}(\tau)$:
\begin{align}
G^{(2)}_s(\tau)
=& \Phi_0^2 \left( \beta^2 \left[ |\psi_s(\tau) |^2+ |\psi_s(0)|^2\right] +\right. \nn & \left.2 \beta  \left[ \psi'_s(\tau) + \psi'_s(0) \right] e^{-\alpha'(0) L} \right),
\label{G2-spon}
\end{align}
where we define \(\psi_s(\tau)\) as the spontaneous emission wavefunction and where the prime index denotes the real part of the quantity.
Note, that the sum \(G^{(2)}_s(\tau) + \Phi_0^2 e^{-2\alpha'(0) L}\) corresponds to the Glauber two-photon correlation function of a field consisting of a coherent pump laser field and a thermal-like field from spontaneous emission in the propagation mode. Unlike the biphoton contribution, no coherence exists between the pump photons and the thermal-like spontaneous photons, resulting in the absence of interference terms in Eq.~\eqref{G2-spon}.

The explicit forms of the functions \(\psi_b(\tau)\) and \(\psi_s(\tau)\) for an open system, with arbitrary relationships between the relaxation constants, are provided in Appendix \ref{appendix}.

The  expressions (\ref{G2biph})--(\ref{G2-spon})  enable the analysis of   the second-order correlation function (\ref{G2_gen}) for light transmitted through a two-level gas under the assumption of a weak input field, considering arbitrary relations between the relaxation parameters of the gas.

\subsection{Nearly closed system regime ($\gamma \rightarrow 0$)}

In the following, we restrict ourselves to the simplest case of a nearly closed system, where spontaneous emission from the upper state \(|2\rangle\) occurs only  to the lower state \(|1\rangle\), and its rate \(\Gamma\) is significantly higher than all other relaxation parameters. Thus, we assume that the atom interacts with the field for a long, but not infinitely long, time \(1/\gamma\). Specifically, we assume that relaxation constants are given by relations   \eqref{closed}. The constant \(\gamma\) accounts for effects such as transit time effects in hot gases, loss of atoms, (e.g., from optical traps), and collisions that cause velocity changes, preventing atoms from interacting with the field due to the Doppler effect and   so on.

Under these assumptions, the biphoton wavefunction \(\psi_{b}(\tau)\) is decomposed into a dynamical term \(\psi_{b}^{(d)}(\tau)\) given by (\ref{psibd}) and the Langevin noise contribution \(\psi_{b}^{(L)}(\tau)\) given by (\ref{psibL}). For ideal closed system, the Fourier transform of the dynamical part simplifies to:
\begin{align}
\psi_{b}^{(d)}(\varpi) &= \frac{4\alpha_0}{\Gamma} \int_0^L d\zeta \, e^{- \frac{1}{2}\alpha(\varpi)(L-\zeta) - \frac{1}{2} \alpha(-\varpi)(L-\zeta) - \alpha(0)\zeta} \nonumber \\
&\times \int dv_z \, W(v_z) \mathcal{L}(\Delta_D + \varpi) \mathcal{L}(-\Delta_D + \varpi) \mathcal{L}(\Delta_D),\label{eq66}
\end{align}
while the Fourier transform of the Langevin noise contribution reads
\begin{align}
\psi_{b}^{(L)}(\varpi) &= -\frac{8\alpha_0}{\Gamma} \int_0^L d\zeta \, e^{- \frac{1}{2}\alpha(\varpi)(L-\zeta) - \frac{1}{2} \alpha(-\varpi)(L-\zeta) - \alpha(0)\zeta} \nonumber \\
&\times \int dv_z \, W(v_z) \mathcal{L}(\Delta_D - \varpi) \mathcal{L}(\Delta_D + \varpi) \nonumber \\
&\times \mathcal{L}(-\Delta_D + \varpi) \mathcal{L}(\Delta_D). \label{eq67}
\end{align}
Here  \( \mathcal{L}(\nu)=(\Gamma/2)/(\Gamma/2-i\nu)\) is a normalized complex Lorentzian function with argument \(\nu\), and
\(\Delta_D=\Delta+kv_z\) is the Doppler-shifted  detuning. For the nearly closed system considered here, the absorption coefficient 
\(\alpha(\varpi)\)
  is   the well-known complex Voigt profile, given by     
\be
\alpha(\varpi)  = \alpha_0 \int dv_z W(v_z)\mathcal{L}(\Delta_D)  .
\label{absorption1}
\ee
\ignore{Using the so-called Faddeeva  function (formerly known as the plasma function)
\be w(\varsigma )={i\over \pi} \int _{-\infty}^\infty dx{  e^{-  x^2}\over \varsigma -x   },
\ee
where \(\varsigma\) is a complex number with \(\Im  \varsigma>0\), the    absorption coefficient \(\alpha(\varpi)\)  can be defined
as
\bea
\alpha(\varpi)&=&\alpha_0\sqrt{\pi} {\eta\over 2  }w(\varsigma),
\eea
where \(\eta= \Gamma/kv_0 \),
\(\varsigma= i\eta/2 + (\Delta+\varpi) /kv_0\).}
In what follows, we will use the detuning-dependent optical depth \( OD \), defined as
\be
OD = \alpha'(0) L.
\ee

The exponential terms in Eqs. \eqref{eq66} and \eqref{eq67}  describe the attenuation of the wavefunction due to absorption, while the integrals over \(\zeta\) and \(v_z\) consider the contributions from different positions in the medium and the velocity distribution, respectively.

Hereafter, we will no longer separate the contributions of the dynamical and Langevin parts. Instead, we will consider the total biphoton wavefunction \( \psi_{b}(\varpi) \), whose expression is remarkably simpler and more symmetric compared to the individual components.
\begin{align}
\psi_{b} (\varpi) &= -   \Gamma {   e^ { - \frac{1}{2}[\alpha(\varpi) +\alpha(-\varpi)]L}  -  e^{ -\alpha(0) L} \over \varpi^2}.
\label{psi_tau}
\end{align}

Using \eqref{appBB}, the spontaneous emission wavefunction \(\psi_{s}(\tau)\) can be written as 
\begin{align}
\psi_{s}(\tau) &=  \frac{4\gamma}{\Gamma^2} \int {d\varpi\over 2\pi} e^{i\varpi\tau}
    { \alpha_0 \left[e^{- \alpha'(\varpi)L  } - e^{- \alpha'(0) L }\right]\over   \alpha'(0) - \alpha'(\varpi)    }  \nn & \times \int dv_z  {W(v_z) }  |  \mathcal{L}(\Delta_D+\varpi)|^2 |  \mathcal{L}(\Delta_D)|^2.
\label{600}
\end{align}

\begin{figure}[h]
 \centerline{\includegraphics[width=90mm,angle=0]{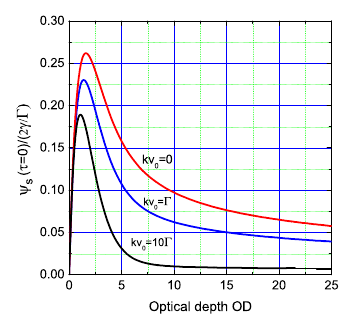}}
 \caption{  Spontaneous emission wavefunction \(\psi_s(\tau )\) for zero delay $\tau=0$ in units of \(2\gamma /\Gamma\) as a function of the optical depth \(OD\)  in the resonance case \(\Delta=0\) for different Doppler widths \(kv_0\) for the case of an almost closed system \(\gamma\ll\Gamma\). }%\vspace*{20mm}
 \label{graph1}
 \end{figure}

Figure \ref{graph1} shows the value of \(\psi_{s}(\tau)\) in units of the small ratio \( {\gamma \over \Gamma/2}\) at zero delay \(\tau=0\) in the resonance case \(\Delta~=~0\) as a function of the optical depth \(OD \) for different Doppler widths \(kv_0\).

It is noteworthy that, in the case of a perfectly closed system, the spontaneous emission term \(\psi_s(\tau) \) is zero such that $G^{(2)}(\tau) = G^{(2)}_b(\tau)$.

\subsubsection*{Low \(OD\) regime}

Let's consider the second-order correlation function in more detail for several specific cases. First, we will analyze the case of small $OD$ when
 \begin{align}
G^{(2)}_b(\tau)
\approx& \Phi_0^2   \left|  1-\alpha(0) L  + \beta \psi_{0}(\tau) \right|^2,
\label{G2-bi-weak}
\end{align}
where
\be
\psi_{0}(\tau)=
 -      \alpha_0 L \int dv_z W(v_z) {  e^{  - (\Gamma/2-i\Delta_D)  \tau  } \over (1-2i\Delta_D/\Gamma)^2  }
 \label{weak-D}
\ee
is the biphoton wavefunction $\psi_{b}(\tau)$ in the limit of small OD. 
 In absence of the Doppler broadening, we recover the result of   \cite{Mah18,scatt} obtained using scattering theory:
 \be
\psi_{0}(\tau)=
 -        \alpha_0 L {  e^{  - (\Gamma/2-i\Delta )  \tau  } \over (1-2i\Delta/\Gamma )^2  } .
 \label{weak-N}
\ee
It is important to note that, in this approximation, the Langevin fluctuations fully account for the result in Eqs.~(\ref{weak-D})--(\ref{weak-N}), as the integral over frequency of the dynamic term, derived in the linear approximation with respect to optical depth, is precisely zero. In contrast, in case of large optical depth (see next subsubsection), both Langevin and dynamical part have comparable contributions.

\begin{figure}[h]
 \centerline{\includegraphics[width=90mm,angle=0]{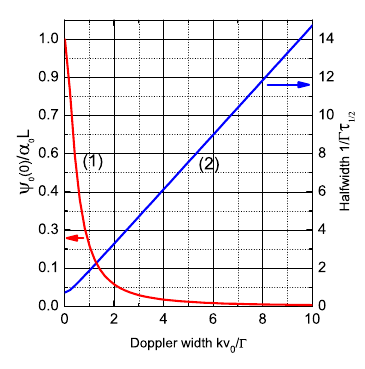}}
  
 \caption{Biphoton wavefunction value at zero delay, \(\tau = 0\), \(\psi_0(0)\) (curve 1), and the effective spectral width, \(1/\tau_{1/2}\) in units of $\Gamma$ (curve 2), where \(\tau_{1/2}\) is the temporal width (half-width at half-maximum) of   \(\psi_0(\tau)\), as a function of Doppler width $k v_0$ for exact resonance, \(\Delta = 0\).}%\vspace*{20mm}
 \label{graph2}
 \end{figure}

The effect of Doppler broadening on the biphoton wavefunction \(\psi_{b}(\tau)\) can be attributed to two primary factors. Firstly, the characteristic width as a function of detuning is determined by the width of the Voigt absorption profile which corresponds to the homogeneous linewidth \(\Gamma/2\)  in the absence of Doppler broadening and matches the Doppler width \(kv_0\)  at very large Doppler broadenings. Secondly, the value of the biphoton wavefunction at zero delay, \(\tau=0\), is proportional  to \(\Gamma/kv_0\). Consequently, for the same widths, the value of the biphoton wavefunction at zero delay, \(\tau=0\), is higher when the contribution of inhomogeneous broadening is smaller. These properties are illustrated in Fig.~\ref{graph2}, which shows the value of the biphoton wavefunction at zero delay, \(\tau=0\), \(\psi_0(0)\), and the effective spectral width, \(1/\tau_{1/2}\), where \(\tau_{1/2}\) is the temporal width (half-width at half-maximum) of   \(\psi_0(\tau)\), as a function of Doppler width $k v_0$ for exact resonance, \(\Delta = 0\).

To characterize the quantum behavior of light, the normalized second-order correlation function \( g^{(2)}(\tau) \) is used. As shown in the Appendix \ref{appenC}, for the case of weak absorption, it suffices to approximate the first-order correlation function as \( G^{(1)}(0) = \Phi_0(1 - \alpha'(0)L) \). Then \( g^{(2)}(\tau) \) can be written as
\begin{equation}
g^{(2)}(\tau) = \frac{G^{(2)}(\tau)}{G^{(1)}(0)^2} \approx 1 + 2 \beta \psi_{0}'(\tau) 
\end{equation}
and we observe the emergence of deviation from Poissonian statistics ($g^{(2)}(0)\neq1$). However, in the regime of low $OD$ and a small parameter \( \beta \), the deviation of \( g^{(2)}(0) \) from unity is very small. Therefore, it is of interest to consider the case of an optically dense medium, which will be addressed in the following subsubsection.

\subsubsection*{Two-photon correlation function for arbitrary OD}

As shown in Eq.~(\ref{G2biph}), the contributions to the correlation function \( G^{(2)}(\tau) \) from the pump field photons and the biphotons add coherently. This coherent addition means their amplitudes   interfere, resulting in   constructive or destructive interference in the overall correlation function. At a specific optical depth \( OD = OD_a \), these contributions can completely cancel each other out at zero delay \(\tau\). Consequently, the correlation function \( G^{(2)}(0) \) reaches its minimum value.

For an ideal closed two-level system, where the decay rate \(\gamma\) is zero, \( g^{(2)}(\tau) \) takes the form:
\begin{align}
	g^{(2)}(\tau)
	=&    {\left| e^{-\alpha(0) L} + \beta \psi_b(\tau) \right|^2\over  e^{-2\alpha'(0) L}}.
	\label{G2biph1}
\end{align}
 To find \( OD_a \), we need the biphoton wavefunction at zero delay \(\tau=0\):
 \begin{align}
\psi_b(0) = - e^{-\alpha(0) L} \int \frac{d\varpi}{2\pi} \frac{\Gamma}{\varpi^2} \left[ e^{\delta\alpha(\varpi) L} - 1 \right],
\label{psi0}
\end{align}
where \(\delta\alpha(\varpi) = \alpha(0) - \frac{1}{2}[\alpha(\varpi) + \alpha(-\varpi)]\).

\begin{figure}[h]
 \centerline{\includegraphics[width=95mm,angle=0]{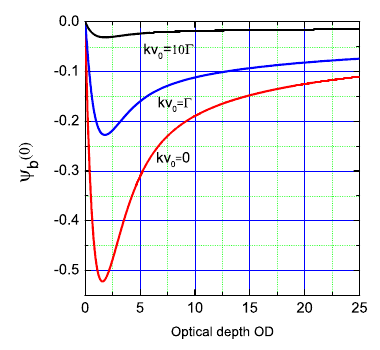}}
\caption{Biphoton wavefunction \(\psi_b(0)\) at zero delay \(\tau = 0\) as a function of the optical depth \(OD\) in the resonant case \(\Delta = 0\) for different Doppler widths \(k v_0 \). }
 \label{graph3}
 \end{figure}

The dependence of \(\psi_{b}(0)\) on optical depth $OD$ is illustrated in Fig.~\ref{graph3} for resonant pumping (\(\Delta=0\)) at various Doppler widths \(kv_0\). Recall that, in the resonant case, the optical depth $OD$ is determined by the absorption coefficient at the center of the Voigt profile. Therefore, for the same optical depth but different Doppler widths, the number of atoms that is required for reaching \(OD_a\) varies. As  seen in Fig.~\ref{graph3},   the maximum absolute value of \( \psi_{b}(0) \) decreases as    \( kv_0/\Gamma  \) increases.

\begin{figure}[t]
	\centerline{\includegraphics[width=71mm,angle=0]{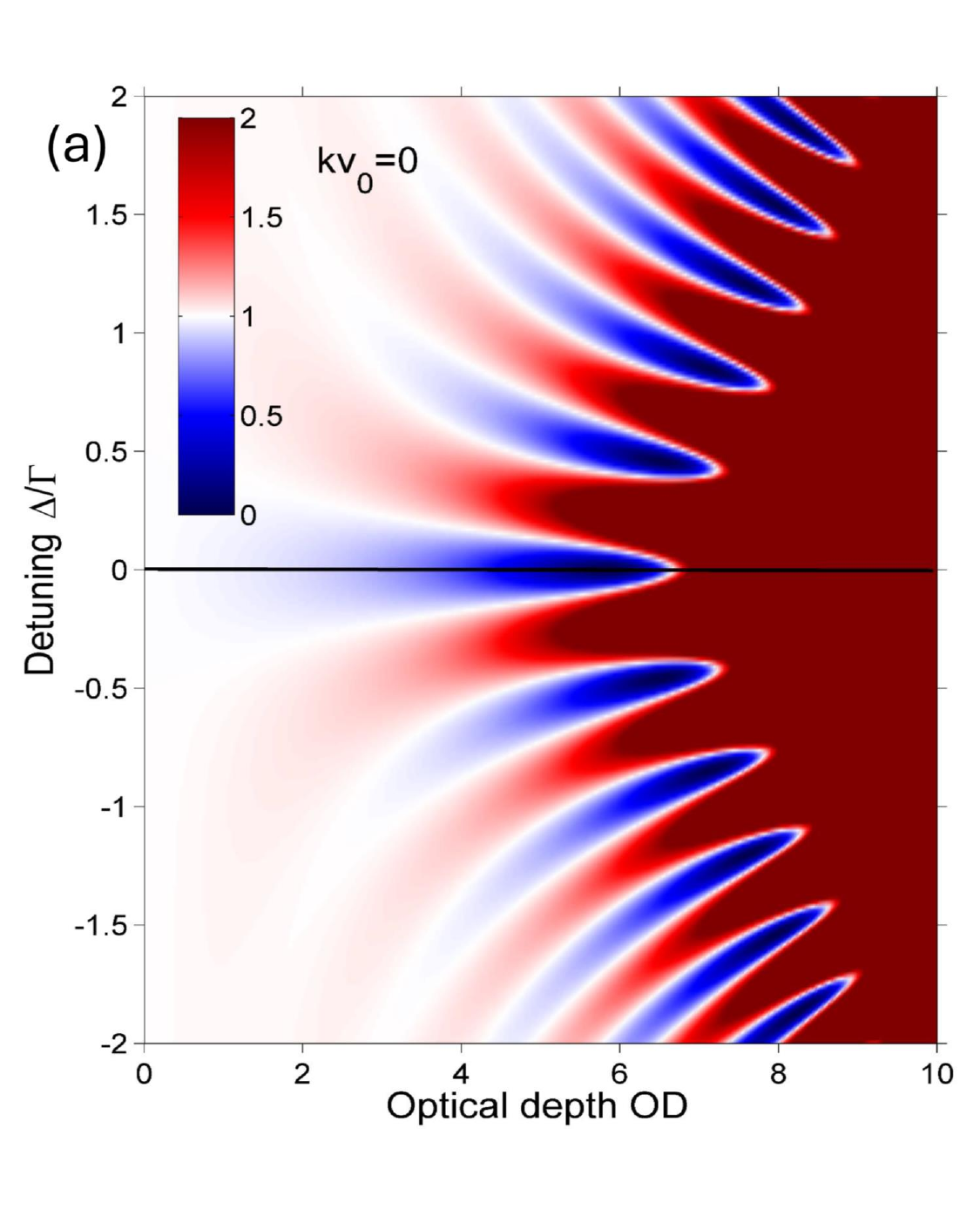}}\vspace*{-4mm}
	\centerline{\includegraphics[width=71mm,angle=0]{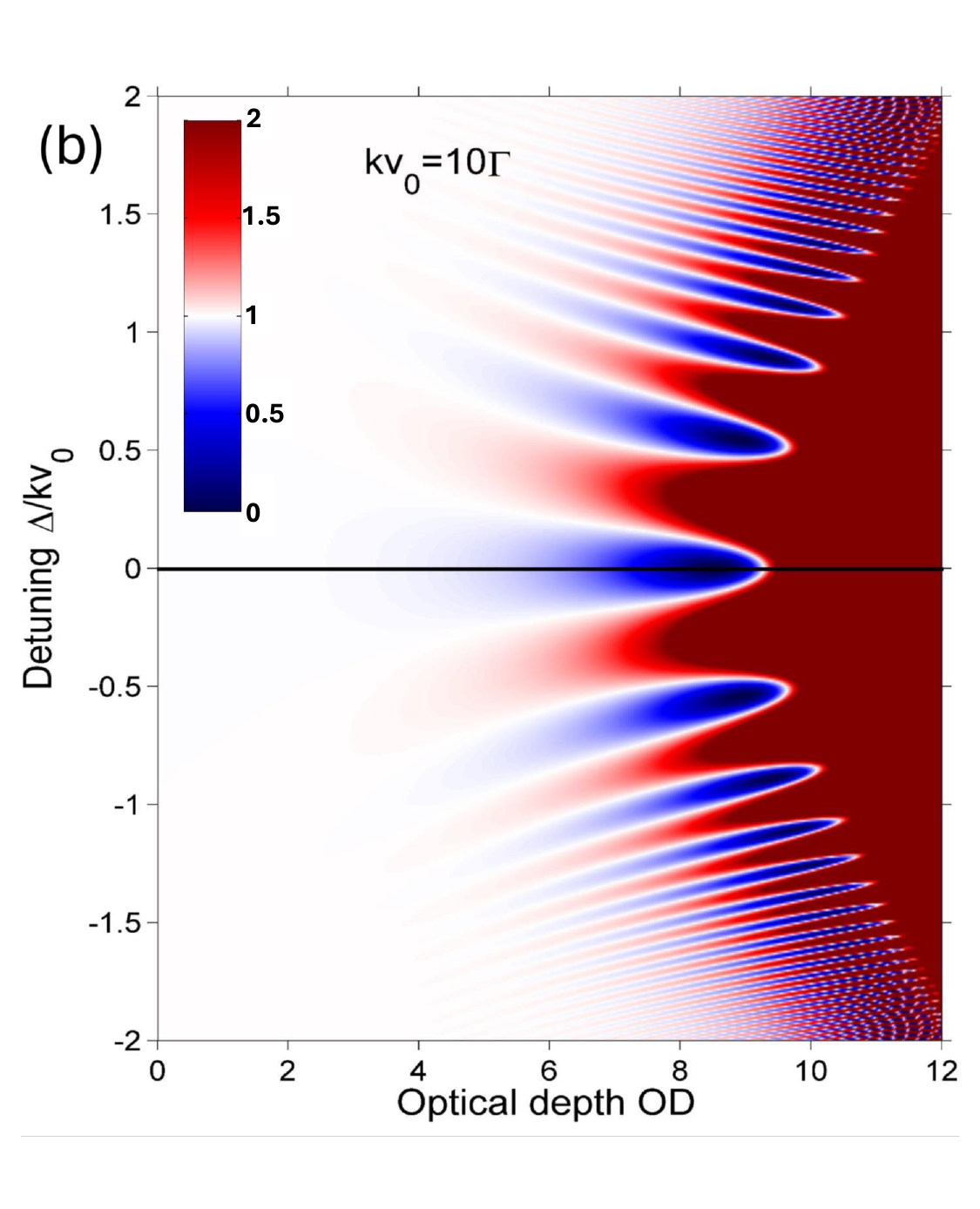}}

	\caption{Density plot of \( g^{(2)}(0) \) as a function of optical depth \( OD \) and detuning \( \Delta \) for \( \beta = 10^{-2} \): (a) No Doppler broadening (\( kv_0 = 0 \)), and (b) Doppler-broadened line with \( kv_0 = 10\Gamma \). For better visibility in the most relevant range, \( 0 < g^{(2)}(0) < 2 \), values of \( g^{(2)}(0) \) greater than 2 are set to \( g^{(2)}(0) = 2 \). Note that the detuning scaling factors differ between panels (a) and (b).}
	 
	\label{graph5}
\end{figure}

Figure \ref{graph5} shows the density plot of \( g^{(2)}(0) \) as a function of optical depth  \(OD\)  and detuning  \(\Delta\), both (a) in the absence of Doppler broadening and (b) for \( kv_0 = 10\Gamma \).

As seen from this plot, there exists a set of points \((OD, \Delta)\)  where complete antibunching is predicted (dark-blue areas). At each of these points, \(\psi_b(0)\) satisfies the equation \(\beta\psi_b(0) = -e^{-\alpha(0)L}\). Since \(\beta \ll 1\), we can use the asymptotic value of \(\psi_b(0)\), obtained in Appendix \ref{appenD} for large optical depths.

In the case of cold atoms ($kv_0  \approx 0$), the asymptotic value of \(\psi_b(0)\) is given by:
\be
\psi_b(0) \approx -\frac{1}{\sqrt{\pi \alpha_0 L \left(1 - \frac{2i\Delta}{\Gamma}\right)}}.
\ee
Using this expression, the antibunching optical depth (\(OD_a\)) at exact resonance (\(\Delta = 0\)) is approximately \(OD_a \approx 6.08\) for \(\beta = 10^{-2}\). With the exact expression for \(\psi_b(0)\) from Eq.~(\ref{psi0}), the solution is \(OD_a \approx 5.88\). For \(\beta = 10^{-3}\), the approximate solution gives \(OD_a \approx 8.55\), while the exact solution is \(OD_a \approx 8.44\).

To find the antibunching points for \(\Delta > 0\), we solve the following nonlinear system of equations:
\begin{align}
	\beta \exp(OD) &= \sqrt{\pi OD} \left(1 + \frac{4\Delta^2}{\Gamma^2}\right)^{3/4}, \\
	OD \frac{2\Delta}{\Gamma} &= 2\pi n - \arctan\left(\frac{2\Delta/\Gamma}{1 + \sqrt{1 + \frac{4\Delta^2}{\Gamma^2}}}\right),
\end{align}
where \(n = 1, 2, \ldots\). The     solutions for $n=1$ and $n=2$  are:
 \(OD_a^{(1)} = 6.56, \Delta_a^{(1)} = 0.450\Gamma\)  and  \(OD_a^{(2)} = 7.16, \Delta_a^{(2)} = 0.841\Gamma.
\)
For negative detuning, antibunching takes place for the same  $OD_a$ when $\Delta=-\Delta_a$.

\begin{figure}[h]
 \centerline{\includegraphics[width=90mm,angle=0]{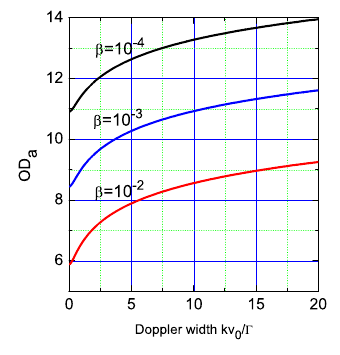}}
\caption{The antibunching optical depth  \(OD_{a}\)  as a function of the Doppler width \(kv_0\) for different values of \(\beta\).}
 \label{antibunch}
 \end{figure}

The influence of Doppler broadening on the optical density required to obtain antibunching,  \(OD_a\), is illustrated in Fig.~\ref{antibunch}, which shows \(OD_a\) as a function of the Doppler width \(kv_0\) for different values of \(\beta\) under resonant pumping. As seen in the figure, \(OD_a\) strongly depends on the type of broadening: for the same absorption line width, achieving antibunching requires a larger \(OD\) for a predominantly Doppler-broadened line ($kv_0\gg\Gamma$) compared to a homogeneously broadened one ($kv_0\ll\Gamma$).

\begin{figure}[h]
 \centerline{\includegraphics[width=90mm,angle=0]{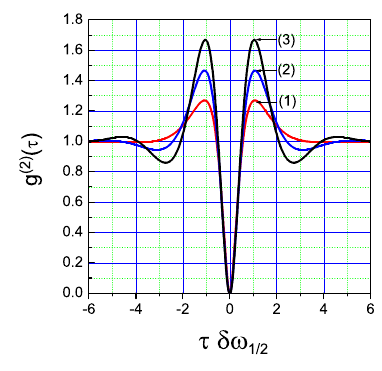}}
\caption{Dependence of the normalized correlation function \(g^{(2)} \) on the normalized time delay \(\tau\delta\omega_{1/2}\) for resonant pumping (\(\Delta=0\)) of the ideal closed system at various Doppler widths: (1) \(kv_0=0, \delta \omega_{1/2}=0.5\Gamma\); (2) \(kv_0=\Gamma, \delta \omega_{1/2}=1.12\Gamma\); (3) \(kv_0=10\Gamma, \delta \omega_{1/2}= 8.60\Gamma\). The optical depth of the medium is chosen to be equal to   \(OD_a\) for \(\beta=10^{-2}\).}
 \label{tau}
 \end{figure}

Figure \ref{tau} shows the dependence of the normalized correlation function \(g^{(2)}\) on the time delay \(\tau\) for resonant pumping (\(\Delta=0\)) of an ideal closed system at various Doppler broadenings \(kv_0\). The optical depth in each case is chosen to be equal to   \(OD_a\) for \(\beta=10^{-2}\). The time delay \(\tau\) is normalized by the factor \(\delta\omega_{1/2}\), which is the half-width at half-maximum (HWHM) of the Voigt profile   for a given \(kv_0\) and $\Gamma$. For the values used in this figure, \(kv_0=0\), \(\Gamma\), and \(10\Gamma\), the corresponding HWHMs are \(\delta\omega_{1/2}=0.5\Gamma\), \(\approx 1.12\Gamma\), and \(\approx 8.60\Gamma\), respectively. As shown, the characteristic width of the function \(g^{(2)}(\tau)\)  is fully determined by the HWHM of the absorption line, regardless of the factors contributing to the line broadening.

\subsection{Open system $\gamma\neq 0$}

In real systems, \(\gamma\) is always non-zero resulting in some contribution to $G^{(2}(\tau)$ from spontaneous emission described by the function $\psi_s(\tau)$ given by Eq.~(\ref{600}). Here, we consider the case $\gamma\ll\Gamma$, which yields   a lower-bound to the minimum value of antibunching that can be reached. 
Let us consider   resonant pumping (\(\Delta=0\)). The condition for maximal  antibunching is \( e^{-OD_a}=\beta|\psi_b(0)|\), for which we get  
 \begin{align} &G^{(2)} (0) =G^{(2)}_s (0)= \Phi_0^2 [ \beta^2   2 \psi_s(0) ^2 + 4 \beta   \psi _s(0)  e^{-OD_a}]= \nn& 2\Phi_0^2\beta^2     [ \psi_s(0) ^2 + 2  \psi _s(0) |\psi_b(0)|]\approx 4\Phi_0^2\beta^2         \psi _s(0) |\psi_b(0)|\nonumber.\end{align} According to Eq.~(\ref{G1}),  \(G^{(1)} (0)\) for the antibunching condition reads \begin{align}G^{(1)} (0)&=\Phi_0\left[ e^{-OD_a}
    +  \beta     \psi_s   ( \tau   )\right]\nn& =\Phi_0\left[  \beta|\psi_b(0)|
    +  \beta     \psi_s   ( \tau   )\right]\approx \Phi_0  \beta|\psi_b(0)|.\nonumber\end{align} As a result, in the case of an open system, the transmitted light does not reach perfect antibunching at \(OD = OD_a\), and the value of the normalized second-order correlation function \(g^{(2)}(0)\) at this \(OD\) is given by:
    \be
     g^{(2)} (0)=4{\psi _s(0)\over |\psi_b(0)|}.
     \ee

Using the asymptotic expressions Eqs.~(\ref{psi_tau_app2}) and Eqs.~(\ref{phi_tau_app5})   for \(\psi _s(0)\) and \(\psi_b(0)\)   in the absence of Doppler broadening, we obtain 
\begin{align}
  g^{(2)}(0) &=   4   \frac{\gamma}{\Gamma}    .
\end{align}

\begin{figure}[h]
 \centerline{\includegraphics[width=80mm,angle=0]{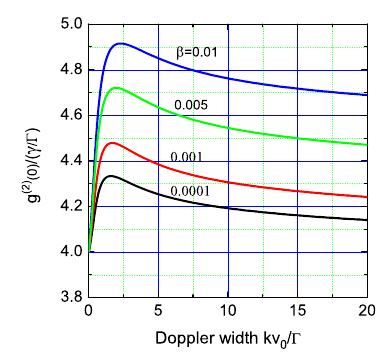}}
\caption{Dependence of \(g^{(2)}(0)\) in units of the ratio \(\gamma/\Gamma\) on Doppler broadenings \(kv_0\) for different values of the parameter \(\beta\)}
 \label{g2}
 \end{figure}

The influence of Doppler broadening on the minimal possible correlation \(g^{(2)}(0)\) is illustrated in Fig.~\ref{g2}, which shows the dependence of \(g^{(2)}(0)\) in units of the ratio \(\gamma/\Gamma\) on Doppler width \(kv_0\) for different values of the parameter \(\beta\). We observe that the minimal value of \(g^{(2)}(0)\) depends very little on \(kv_0\) and \(\beta\), and over a wide range of parameters, it remains approximately \(( 4...5)  \gamma/\Gamma \).

 \section{Comparison with experimental data} 

\label{Sec6}

\begin{figure}[h]
	\centerline{\includegraphics[width=92mm,angle=0]{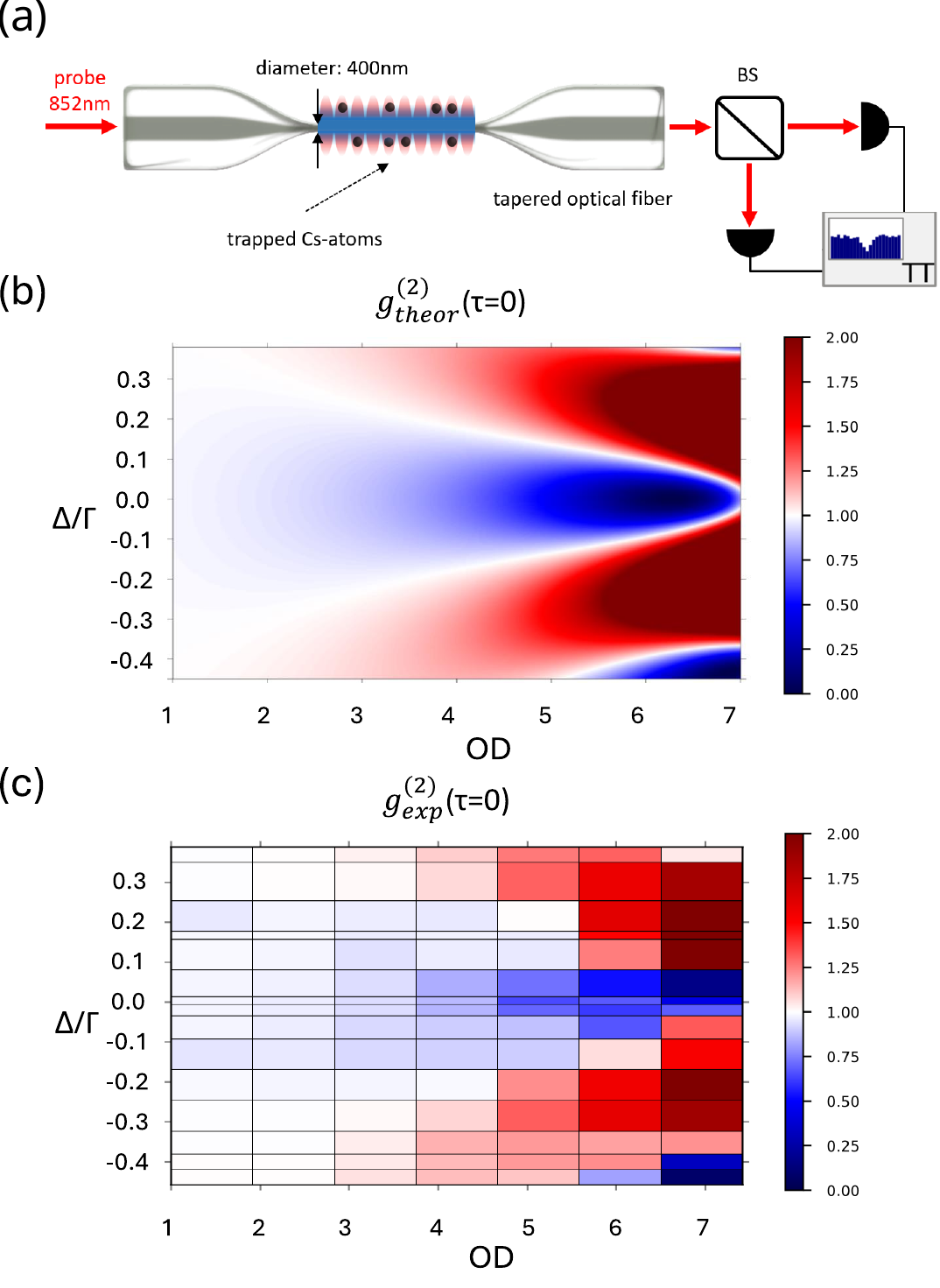}}
	\caption{ (a) Schematic of the experimental set-up. A near-monochromatic laser field is transmitted through an ensemble of laser-cooled cesium atoms, optically trapped and interfaced using the evanescent field of an optical nanofiber.  The photon statistics is measured using a Hanbury-Brown and Twiss setup.   Model predictions (b)  and experimental data (c)   for \(g^{(2)}(\tau=0)\) as a function of optical depth  OD  and detuning \(\Delta\). The model is calculated for the experimental value of \(\beta = 0.007\) and corresponds to a smaller range of OD and detuning compared to Fig.~\ref{graph5}a.   }
	\label{fig:comparison}
\end{figure}

Here we compare the model predictions with experimental results obtained from a waveguide QED platform, where cold atoms are coupled to an optical nanofiber~\cite{Vetsch10}.  Specifically, an ensemble of laser-cooled cesium (Cs) atoms is trapped in two diametrically opposed 1D dipole trap arrays, positioned approximately 250 nm above and below the nanofiber, which is realized as the waist of a tapered optical fiber (see Fig.~\ref{fig:comparison}a).
The probe light, near resonant with the $D_2$-line (natural linewidth \(\Gamma = 2\pi \times 5.2\) MHz), is launched into the nanofiber wave\-guide and couples evanescently to the atoms with a coupling constant of \(\beta = 0.007 \pm 0.002\) \cite{Cor23}. The photon statistics is measured using a Hanbury-Brown and Twiss setup for various optical depths  $OD$  and laser-atom detunings, \(\Delta = \omega_{12} - \omega_p\), where \(\omega_{12}\) is the \(6S_{1/2}, F=4 \rightarrow 6P_{3/2}, F'=5\) transition frequency. The detuning \(\Delta\) is varied using an acousto-optic modulator, and the optical depth is adjusted by varying the   time interval  between loading the nanofiber-based trap and probing the atoms.

The photon statistics of the light transmitted through the atomic ensemble in these experiments have been characterized both on resonance~\cite{Pra20} and under detuned excitation~\cite{Cor23}. Figure~\ref{fig:comparison}c presents additional, previously unpublished results of measurements of the second-order correlation function \(g^{(2)}(\tau)\) at \(\tau=0\) as a 2D plot, with detuning \(\Delta\) and optical depth \(OD\) as variables. Comparing these with the model predictions in Fig.~\ref{fig:comparison}b, we observe   good quantitative agreement,   providing a robust validation of the model and highlighting its ability to accurately describe the photon statistics in the system.

\section{Conclusions}
 
In this paper, we have presented a theoretical analysis of the quantum behavior of nearly monochromatic light as it propagates through an atomic gas medium consisting of two-level atoms, utilizing the Heisenberg-Langevin equation method. Our study has primarily focused on investigating the antibunching behavior of the transmitted light, created by the interference between the coherent pump field and the generated biphotons. We also analyzed the contribution of photons spontaneously emitted into the spatial mode of the propagating pump laser radiation to the second-order correlation function \( g^{(2)}(\tau) \).

We have demonstrated that in the specific case of a closed system without Doppler broadening, the derived expressions reproduce known dependencies recently obtained using scattering theory, establishing conditions for complete photon antibunching. We compared our theoretical predictions with experimental results obtained using a waveguide QED platform with cold atoms coupled to an optical nanofiber and obtained the good agreement between the experimental data and our model predictions. 

We emphasize that, since we are using a continuous medium model, our approach does not impose limitations on the number of atoms, as is the case in existing models. Moreover, the Heisenberg-Langevin approach enables the analysis of significantly more complex systems that are challenging to study using conventional scattering theory methods. Specifically, we analyzed the influence of Doppler broadening on antibunching conditions in a closed system, the role of pump field detuning, and the consequences of finite interaction time for an almost closed system.

The model can be readily adapted to other inhomogeneously broadened systems, such as doped crystals, systems with additional decoherence like quantum dots, and high-energy cases of x-ray excitations of nuclear M\"o\ss{}bauer transitions.

Future work could extend this analysis to more complex many-level systems and explore other characteristics of quantum light, such as squeezing, three-photon and higher-order correlations, and strategies to minimize the impact of physical and technical factors on \( g^{(2)}(\tau) \).  These efforts will provide useful insights for improving quantum light sources in practical applications. 

\begin{acknowledgments}
 
L. P. acknowledges support from the Einstein Foundation (Einstein Research Unit on Quantum Devices).   M.S. acknowledges support by Young
Researcher Grant MajorSuperQ MSCA. We acknowledge funding by the Alexander von Humboldt Foundation in the framework of the Alexander von Humboldt Professorship endowed by the Federal Ministry of Education and Research, as well as direct funding by the Federal Ministry of Education and Research (Collaborative Research Project NetiQueT).

\end{acknowledgments}

\begin{widetext}
\appendix

\section{ Correlation relations }\label{appendix1}

Here, we derive the correlation relations for the linear \(\hat{B}(\varpi, z)\) and nonlinear \(\hat{C}(\varpi, z)\) contributions (see Eqs. (\ref{zero_order}) and (\ref{first_order})) to the solution
\be
\hat{a}(\varpi, z) = \sqrt{2\pi} \hat{a}_{p}(z) \delta(\varpi) + \hat{B}(\varpi, z) + \hat{C}(-\varpi, z)^\dag
\label{solApp}
 \ee
 for the annihilation operator \(\hat{a}(\varpi, z)\). These contributions are essential for calculating the mean values of the operators of interest.

To apply the diffusion coefficients described in Eq. (\ref{Diff-matrix}), we require the operators \( \hat{s}_{ij}\). We will seek them within the weak saturation approximation $S\ll 1$, where the saturation parameter \(S\) is defined by Eq. (\ref{sat_par}). Specifically, we need the stationary solutions \(\hat{\sigma}^{(0)}_{ij}\) from Eqs. (\ref{eq12})--(\ref{eq22}) when \(\hat{f}_{\alpha}=0\) and $\hat{a}(t,z)=\hat{a}_{p}(z)$.
 They read:
\begin{align}
  \hat{s} _{11}(z)   &= 1 - g^2 \hat{a}_{p}^\dag(z) \hat{a}_{p}(z) \frac{2 \gamma_{12}}{\gamma_1 G_{12}(0) G_{21}(0)} \left(1 - \frac{\Gamma}{\gamma_2}\right), \label{sigma01} \\
 \hat{s}_{22}(z)   &= g^2 \hat{a}_{p}^\dag(z) \hat{a}_{p}(z) \frac{2 \gamma_{12}}{\gamma_2 G_{12}(0) G_{21}(0)}, \\
  \hat{s}_{12}(z)  &= i g \frac{\hat{a}_{p}(z)}{G_{12}(0)}, \\
  \hat{s}_{21}(z)   &= -i g \frac{\hat{a}_{p}^\dag(z)}{G_{21}(0)}. \label{sigma02}
\end{align}

The correlation \(\langle \hat{B}(\varpi, \zeta) \hat{B}(\varpi', \zeta')^\dag \rangle_a\) depends on the relation between \(\zeta\) and \(\zeta'\):
\begin{align}
\langle \hat{B}(\varpi, \zeta) \hat{B}(\varpi', \zeta')^\dag \rangle_a &= e^{-\frac{1}{2}\alpha(\varpi)(\zeta - \zeta')} \delta(\varpi - \varpi') \quad \text{for} \quad \zeta > \zeta', \\
\langle \hat{B}(\varpi, \zeta) \hat{B}(\varpi', \zeta')^\dag \rangle_a &= e^{-\frac{1}{2}\alpha(\varpi)^*(\zeta' - \zeta)} \delta(\varpi - \varpi') \quad \text{for} \quad \zeta < \zeta'.
\end{align}

The correlation \(\langle \hat{B}(\varpi, L)^\dag \hat{B}(\varpi', L) \rangle_a\) describes incoherent radiation of spontaneous photons in propagation mode. It is not zero only for an open system, when \(\gamma_{12}>\gamma_2/2\):
\begin{align}
\langle \hat{B}(\varpi, L)^\dag \hat{B}(\varpi', L) \rangle_a &= \beta \hat{a}_{p}^\dag(0) \hat{a}_{p}(0) \psi_{s}(\varpi) \delta(\varpi - \varpi'),
\end{align}
where
\begin{align}
\psi_{s}(\varpi)
= & \left({2\gamma_{12}\over\gamma_2} -1\right)   \alpha_0 { e^{- \alpha'(\varpi)  } - e^{- \alpha'(0) L }\over   \alpha'(0) - \alpha'(\varpi)    }   \int dv_z  {W(v_z) } \frac{ \Gamma \gamma_{12}^2 }{ |G_{12} (\varpi )|^2|G_{12} (0 )|^2}.
\label{appBB}
\end{align}

The correlation between the linear \(\hat{B}(\varpi, L)\) and nonlinear \(\hat{C}(\varpi, L)\) contributions to the solution for annihilation operators, which describes biphoton creation, is given by:
\be
\langle \hat{B}(-\varpi, L) \hat{C}(-\varpi', L)^\dag \rangle = \beta \hat{a}_{p}(0) \hat{a}_{p}(0) \psi_{b} (\varpi)\delta(\varpi - \varpi').
\label{appBC}
\ee
Here, the so-called biphoton wavefunction \(\psi_{b} (\varpi)\) is the sum of the dynamical contribution \(\psi_{b}^{(d)}(\varpi)\) and the contribution \(\psi_{b}^{(L)}(\varpi)\) due to Langevin fluctuations:
\begin{align}
\psi_{b} (\varpi)=\psi_{b}^{(d)}(\varpi)+
\psi_{b}^{(L)}(\varpi),\end{align}
\be
\psi_{b}^{(d)}(\varpi) = \alpha_0 \frac{e^{-\delta\alpha(\varpi)L} - 1}{\delta\alpha(\varpi)} e^{-\alpha(0)L} \int dv_z W(v_z) \frac{\Gamma\gamma_{12}[G_{21}(\varpi) + G_{12}(0)]}{2G_0(\varpi) G_{12}(0) G_{12}(\varpi) G_{21}(\varpi)},
\label{psibd}
\ee
\begin{align}
\psi_{b}^{(L)}(\varpi)&= -\alpha_0 {e^{-\delta\alpha(\varpi)L}-1\over \delta\alpha(\varpi) } e^{-\alpha(0)L}  \int dv_z W(v_z)     \frac{  \gamma_{12} \Gamma   }{2 G_{12}(-\varpi )G_{12}(\varpi)}
 \left[{1\over G_0(\varpi)} \left( {  2\gamma_{12} \over G_{21}(\varpi)}  +  { \gamma_2 \over G_{12}(0 ) }   \right)    +   {  \gamma_1+\Gamma -\gamma_2 \over     G_{11}(\varpi)G_{12}(0 ) }       \right] ,
\label{psibL}
\end{align}
where
\be
\delta\alpha(\varpi)=\alpha(0)-{1\over 2}\left[\alpha(\varpi)+\alpha(-\varpi)\right].
\ee

 \section{Derivation of the  Glauber two-photon correlation function}\label{appendix}

 The Glauber two-photon correlation function of the output field is defined as:
\be
G^{(2)}(\tau) = \langle \hat{a}^\dag(t,L)\hat{a}^\dag(t+\tau,L) \hat{a}(t+\tau,L)\hat{a}(t,L) \rangle.
\ee
In the Fourier space, this can be expressed as:
\be
G^{(2)}(\tau) = \int \frac{d\varpi_1}{\sqrt{2\pi}} \frac{d\varpi_2}{\sqrt{2\pi}} \frac{d\varpi_3}{\sqrt{2\pi}} \frac{d\varpi_4}{\sqrt{2\pi}} e^{+i\varpi_1 t + i\varpi_2(t+\tau) - i\varpi_3(t+\tau) - i\varpi_4 t} \langle \hat{a}^\dag(\varpi_1, L) \hat{a}^\dag(\varpi_2, L) \hat{a}(\varpi_3, L) \hat{a}(\varpi_4, L) \rangle.
\ee

As discussed in Subsection \ref{subsec_Gen_exp}, the solution (\ref{solution}) for the annihilation operator   can be used to calculate the second-order correlation function only if the field entering the medium is weak enough, allowing us to neglect terms in \(G^{(2)}(\tau)\) that are proportional to the sixth and higher orders of the entering flux \(\Phi_0\). Keeping only the terms that contribute proportionally to \(\Phi_0^2\) and are non-zero after averaging using the correlation behavior of the solution, we obtain:
\begin{align}
G^{(2)}(\tau) &= \langle \hat{a}_{p}(L)^\dag \hat{a}_{p}(L)^\dag \hat{a}_{p}(L) \hat{a}_{p}(L) \rangle \\
&\quad + \frac{1}{2\pi} \int d\varpi_1 d\varpi_2 e^{i(\varpi_1 - \varpi_2)t + i\varpi_1 \tau} \langle \hat{a}_{p}(L)^\dag \hat{B}(\varpi_1, L)^\dag \hat{a}_{p}(L) \hat{B}(\varpi_2, L) \rangle \\
&\quad + \frac{1}{2\pi} \int d\varpi_1 d\varpi_2 e^{i(\varpi_1 - \varpi_2)(t + \tau)} \langle \hat{a}_{p}(L)^\dag \hat{B}(\varpi_1, L)^\dag \hat{B}(\varpi_2, L) \hat{a}_{p}(L) \rangle \\
&\quad + \frac{1}{2\pi} \int d\varpi_1 d\varpi_2 e^{i(\varpi_1 - \varpi_2)t - i\varpi_1 \tau} \langle \hat{B}(\varpi_1, L)^\dag \hat{a}_{p}(L)^\dag \hat{B}(\varpi_2, L) \hat{a}_{p}(L) \rangle \\
&\quad + \frac{1}{2\pi} \int d\varpi_1 d\varpi_2 e^{i(\varpi_1 - \varpi_2)t} \langle \hat{a}_{p}(L)^\dag \hat{B}(\varpi_1, L)^\dag \hat{B}(\varpi_2, L) \hat{a}_{p}(L) \rangle \\
&\quad + \frac{1}{(2\pi)^2} \int d\varpi_1 d\varpi_2 d\varpi_3 d\varpi_4 e^{i(\varpi_1 + \varpi_2 - \varpi_3 - \varpi_4)t + i(\varpi_2 - \varpi_3) \tau} \langle \hat{B}(\varpi_1, L)^\dag \hat{B}(\varpi_2, L)^\dag \hat{B}(\varpi_3, L) \hat{B}(\varpi_4, L) \rangle \\
&\quad + \frac{1}{2\pi} \int d\varpi_1 d\varpi_2 e^{i(\varpi_1 + \varpi_2)t + i\varpi_2 \tau - i\varpi_1 \tau} \langle \hat{C}(-\varpi_1, L) \hat{B}(\varpi_2, L)^\dag \hat{a}_{p}(L) \hat{a}_{p}(L) \rangle \\
&\quad + \frac{1}{2\pi} \int d\varpi_1 d\varpi_2 e^{-i(\varpi_1 + \varpi_2)t - i\varpi_2 \tau} \langle \hat{a}_{p}(L)^\dag \hat{a}_{p}(L)^\dag \hat{B}(\varpi_1, L) \hat{C}(-\varpi_2, L)^\dag \rangle \\
&\quad + \frac{1}{(2\pi)^2} \int d\varpi_1 d\varpi_2 d\varpi_3 d\varpi_4 e^{i(\varpi_1 + \varpi_2 - \varpi_3 - \varpi_4)t + i(\varpi_2 - \varpi_3) \tau} \langle \hat{C}(-\varpi_1, L) \hat{B}(\varpi_2, L)^\dag \hat{B}(\varpi_3, L) \hat{C}(-\varpi_4, L)^\dag \rangle.
\end{align}

Recall, that \(\hat{a}_{p}(L)\) is the   pump field annihilation operator \eqref{aP} and \(\hat{B}(\varpi, L)\) and \(\hat{C}(\varpi, L)\) are the linear and nonlinear contributions to the solution \eqref{solution} for the annihilation operator  \(\hat{a} (\varpi,z)\) at position \(z=L\), respectively. The correlation function \(G^{(2)}(\tau)\) includes contributions from various combinations of these operators, integrated over the frequency variables \(\varpi_i\).
Taking into account that Langevin fluctuations are Gaussian processes, the pump depletion law is    \(\hat{a}_{p}(L)=\hat{a}_{p}(0)e^{-{1\over 2}\alpha(0)L}\)  and,  using correlations (\ref{appBB}) and (\ref{appBC}), we obtain
 \begin{align}
G^{(2)}(\tau)
& =
\Phi_0^2\left( \left|e^{-\alpha(0)L }+ \beta   \psi_{b} (\tau) \right|^2+ \beta  e^{-\alpha(0)'L } [ \psi_{s}^*(\tau)+
 2  \psi_{s}(0)+
         \psi_{s}(\tau)]+
  \beta^2   \left[|\psi_{s}(\tau)|^2  + \beta^2  |\psi_{s}(0)|^2 \right] \right),
\label{C2fin}
\end{align}
   where 
   \begin{equation}
 \psi_{s,b}(\tau)=   {1 \over {2\pi} } \int      d\varpi       e^{
-i\varpi   \tau   }             \psi_{s,b}(\varpi).
\end{equation}

\section{Normalized autocorrelation function}
\label{appenC}
 The normalized second-order Glauber autocorrelation function is defined as \( g^{(2)}(\tau) = \frac{G^{(2)}(\tau) }{[G^{(1)}(0) ]^2}   \),
where the first-order autocorrelation function \( G^{(1)}(\tau) \) reads
\be
 G^{(1)}(\tau) =\langle \hat{a}^\dag(t,L) \hat{a}(t+\tau,L) \rangle= \int  { d\varpi_1 \over \sqrt{2\pi} }{ d\varpi_2 \over \sqrt{2\pi} }  \langle \hat{a}(\varpi_1,L)^\dag\hat{a}( \varpi_2 ,L)   \rangle e^{+i[\varpi_1 t- \varpi_2(t+\tau)]}.
 \ee
 Since we are interested in the correlation function \( g^{(2)}(\tau) \) in the case of a weak input field, it is sufficient to know the first-order correlation function in the linear approximation with respect to the input photon flux \(\Phi_0= \langle  \hat{a}_{p} ( 0)^\dag \hat{a}_{p} (0)\rangle\). Using solution   (\ref{solApp}) for the annihilation operator, we obtain the following expression:
\begin{align}
 G^{(1)}(\tau) &  =
   \langle  \hat{a}_{p} ( L)^\dag \hat{a}_{p} ( L)\rangle  + \int  { d\varpi_1 \over \sqrt{2\pi} }{ d\varpi_2 \over \sqrt{2\pi} }
   \left[ \langle \hat{B} (\varpi_1 ,L)^\dag\hat{B} (\varpi_2 ,L)\rangle + \langle \hat{C} (-\varpi_1 ,L)
    \hat{C} (-\varpi_2 ,L)^\dag\rangle \right]  \nn
     =&\Phi_0\left[ e^{-\alpha'(0  ) L}
    +  \beta     \psi_s   ( \tau   )\right],
    \label{G1}
 \end{align}
 where
\( \psi_s   ( \tau ) \) is given by expression (\ref{601}).

For weak absorption
\begin{align}
 G^{(1)}(0)     =&\Phi_0 \left[1 - { \alpha'(0  )\over\alpha_0} OD
    + 4  {\gamma\over \Gamma}OD \int    { d\varpi  \over   2\pi \Gamma }     \kappa_c(\varpi)
  +{16\beta^2  \over  \Gamma^2} \Phi_0 4OD \int    { d\varpi  \over   2\pi  }
  \kappa_0(\varpi) \right].
 \end{align}
 Since \(\gamma\ll \Gamma\) and \(\beta\ll1\) we have
 \begin{align}
 G^{(1)}(0)     =&\Phi_0 \left[1 - { \alpha'(0  )\over\alpha_0} OD\right].
 \end{align}

\section{Asymptotic behavior of the biphoton wavefunction}
\label{appenD}
 Here, we consider the biphoton wavefunction under the assumption that the optical depth \( OD = \alpha(0)L \) is large, meaning \( e^{-OD} \ll 1 \)  even if the pump field is detuned from the line center and the absorption line is Doppler broadened. We consider the biphoton wavefunction \(\psi_b(\tau = 0)\) given by:
\begin{align}
\psi_{b} (0) &= - \int_{-\infty}^{\infty} \frac{d\varpi}{2\pi} \Gamma \frac{e^{- \frac{1}{2}[\alpha(\varpi) + \alpha(-\varpi)]L} - e^{-\alpha(0)L}}{\varpi^2}.
\label{psi_tau_app}
\end{align}

For large \( OD \), we can neglect \( e^{-\alpha(0)L} \), introduce a new variable \( x = \Gamma^2 / (4\varpi^2 + \Gamma^2) \), and obtain:
\begin{align}
\psi_{b} (0) &= -2 \int_{0}^{1} \frac{dx}{2\pi} \frac{1}{\sqrt{x}(1-x)^{3/2}} e^{- \frac{1}{2}[\alpha(\Gamma \sqrt{1/x-1}/2) + \alpha(-\Gamma \sqrt{1/x-1}/2)]L}.
\label{psi_tau_app_s}
\end{align}

Considering that \( W(v_z) = W(-v_z) \) and that  for large \( OD \)  the main contribution to this integral comes from \( x < 1/OD \), we have:
\begin{align}
\psi_{b} (0) &\approx   -\frac{1}{\pi\sqrt{\alpha_0 L(1 - 2i\Delta/\Gamma)}} \int_{0}^{\infty} dy \frac{1}{\sqrt{y}} e^{-y - y^2/8\mu^2},
\label{psi_tau_app26}
\end{align}
where   
\be
\mu = {\Gamma\over kv_0} \left[\frac{\alpha_0 L(1 - 2i\Delta/\Gamma)}{48}\right]^{1/2}.
\ee

After integrating over \( y \), we get:
\begin{align}
\psi_{b} (0) &= -\frac{\eta}{\sqrt{24}\pi} e^{\mu^2} K_{1/4}(\mu^2),
\label{psi_tau_app27}
\end{align}
where \( K_{\nu}(x) \) is the modified Bessel function of the second kind.

In the absence of Doppler broadening (\(kv_0\ll \Gamma\)),
\begin{align}
\psi_{b} (0) &= -\frac{1}{\sqrt{\pi \alpha_0 L(1 - 2i\Delta/\Gamma)}}.
\label{psi_tau_app2}
\end{align}

For a Doppler-broadened line with \( |\mu| \ll 1 \), we have
\be
\psi_{b} (0) = -\frac{1}{24^{1/4} \Gamma(3/4)} \frac{\sqrt{\eta}}{[\alpha_0 L(1 - 2i\Delta/\Gamma)]^{1/4}},
\ee
where \(\Gamma(3/4) \approx 1.225\), with \(\Gamma(x)\) representing the Euler gamma function.
Note the slow decay of \( \psi_{b} (0) \) with \( \alpha_0 L \): \( \propto 1/(\alpha_0 L)^{1/4} \), in contrast to the case without Doppler broadening, where \( \psi_{b} (0) \propto 1/(\alpha_0 L)^{1/2} \). However, this holds until \( OD < kv_0/\Gamma \). For \( OD \gg kv_0/\Gamma \), we observe a faster decay of \( \psi_{b} (0) \propto 1/(\alpha_0 L)^{1/2} \).

\section{Asymptotic behavior of the spontaneous emission wavefunction \(\psi_{s}(0)\)}

Here, we consider the spontaneous emission wavefunction \(\psi_{s}(0)\) under the assumption of large optical depth. We consider an  almost closed system and limit ourselves by the simplest case of resonance and cold atoms where
 \begin{align}
\psi_{s}(0) =&\gamma   \int {d\varpi\over 2\pi} {  1\over  \varpi^2} { e^{- \alpha'(\varpi)L  }     }.
\label{601}
 \end{align}
For large \( OD \), we can neglect \( e^{-\alpha(0)L} \), introduce a new variable \( x = \Gamma^2 / (4\varpi^2 + \Gamma^2) \), and obtain:
\begin{align}
\psi_{s} (0) &= {\gamma\over \Gamma } \int_{0}^{\infty} \frac{dx}{ \pi} \frac{1}{\sqrt{x} } e^{- \alpha_0 L x}={\gamma\over \Gamma }{1\over \sqrt{\pi \alpha_0 L}}.
\label{phi_tau_app5}
\end{align}

\end{widetext}


\begin{thebibliography}{0}%
\makeatletter
\providecommand \@ifxundefined [1]{%
 \@ifx{#1\undefined}
}%
\providecommand \@ifnum [1]{%
 \ifnum #1\expandafter \@firstoftwo
 \else \expandafter \@secondoftwo
 \fi
}%
\providecommand \@ifx [1]{%
 \ifx #1\expandafter \@firstoftwo
 \else \expandafter \@secondoftwo
 \fi
}%
\providecommand \natexlab [1]{#1}%
\providecommand \enquote  [1]{``#1''}%
\providecommand \bibnamefont  [1]{#1}%
\providecommand \bibfnamefont [1]{#1}%
\providecommand \citenamefont [1]{#1}%
\providecommand \href@noop [0]{\@secondoftwo}%
\providecommand \href [0]{\begingroup \@sanitize@url \@href}%
\providecommand \@href[1]{\@@startlink{#1}\@@href}%
\providecommand \@@href[1]{\endgroup#1\@@endlink}%
\providecommand \@sanitize@url [0]{\catcode `\\12\catcode `\$12\catcode `\&12\catcode `\#12\catcode `\^12\catcode `\_12\catcode `\%12\relax}%
\providecommand \@@startlink[1]{}%
\providecommand \@@endlink[0]{}%
\providecommand \url  [0]{\begingroup\@sanitize@url \@url }%
\providecommand \@url [1]{\endgroup\@href {#1}{\urlprefix }}%
\providecommand \urlprefix  [0]{URL }%
\providecommand \Eprint [0]{\href }%
\providecommand \doibase [0]{https://doi.org/}%
\providecommand \selectlanguage [0]{\@gobble}%
\providecommand \bibinfo  [0]{\@secondoftwo}%
\providecommand \bibfield  [0]{\@secondoftwo}%
\providecommand \translation [1]{[#1]}%
\providecommand \BibitemOpen [0]{}%
\providecommand \bibitemStop [0]{}%
\providecommand \bibitemNoStop [0]{.\EOS\space}%
\providecommand \EOS [0]{\spacefactor3000\relax}%
\providecommand \BibitemShut  [1]{\csname bibitem#1\endcsname}%
\let\auto@bib@innerbib\@empty
%</preamble>
\end{thebibliography}%


\begin{thebibliography}{00}
 
\bibitem{Ham10} K. Hammerer, A. S. S\o rensen, and E. S. Polzik, Quantum interface between light and atomic ensembles, Rev. Mod. Phys. \textbf{82}, 1041 (2010).

\bibitem{Alle87} L. Allen and J. H. Eberly, \textit{Optical Resonance and Two-Level Atoms} (Dover Publications, New York, 1987).

\bibitem{Fle05} M. Fleischhauer, A. Imamoglu, and J. P. Marangos, Electromagnetically induced transparency: Optics in coherent media, Rev. Mod. Phys. \textbf{77}, 633 (2005).

\bibitem{Har06} S. Haroche and J. M. Raimond, \textit{Exploring the Quantum: Atoms, Cavities, and Photons} (Oxford University Press, Oxford, 2006).

\bibitem{Kolo07} M. I. Kolobov, ed., \textit{Quantum Imaging} (Springer, New York, 2007).

\bibitem{Bir05} K. M. Birnbaum, A. Boca, R. Miller, A. D. Boozer, T. E. Northup, and H. J. Kimble, Photon blockade in an optical cavity with one trapped atom, Nature (London) \textbf{436}, 87 (2005).

\bibitem{Wal08} D. F. Walls and G. J. Milburn, \textit{Quantum Optics} (Springer, Berlin, 2008).

\bibitem{Dal83} J. Dalibard and S. Reynaud, Correlation signals in resonance fluorescence: Interpretation via photon scattering amplitudes, J. Phys. (Paris) \textbf{44}, 1337 (1983).

\bibitem{Coh98} C. Cohen-Tannoudji, J. Dupont-Roc, and G. Grynberg, \textit{Atom-Photon Interactions: Basic Processes and Applications} (Wiley, New York, 1998).

\bibitem{Gar85} C. W. Gardiner and M. J. Collett, Input and output in damped quantum systems: Quantum stochastic differential equations and the master equation, Phys. Rev. A \textbf{31}, 3761 (1985).

\bibitem{Car09} H. J. Carmichael, \textit{An Open Systems Approach to Quantum Optics} (Springer, Berlin, 2009).

\bibitem{Kol07} P. Kolchin, Electromagnetically-induced-transparency-based paired photon generation, Phys. Rev. A \textbf{75}, 033814 (2007).

\bibitem{Ray07} C. H. Raymond Ooi, Q. Sun, M. S. Zubairy, and M. O. Scully, Correlation of photon pairs from the double Raman amplifier: Generalized analytical quantum Langevin theory, Phys. Rev. A \textbf{75}, 013820 (2007).

\bibitem{Du23} Y. Jiang, Y. Mei, and S. Du, Quantum Langevin theory for two coupled phase-conjugated electromagnetic waves, Phys. Rev. A \textbf{107}, 053703 (2023).

\bibitem{Mah18} S. Mahmoodian, M. \v{C}epulkovskis, S. Das, P. Lodahl, K. Hammerer, and A. S. S\o rensen, Strongly correlated photon transport in waveguide quantum electrodynamics with weakly coupled emitters, Phys. Rev. Lett. \textbf{121}, 143601 (2018).

\bibitem{Pra20} A. S. Prasad, J. Hinney, S. Mahmoodian, K. Hammerer, S. Rind, P. Schneeweiss, A. S. S\o rensen, J. Volz, and A. Rauschenbeutel, Correlating photons using the collective nonlinear response of atoms weakly coupled to an optical mode, Nat. Photonics \textbf{14}, 719 (2020).

\bibitem{Hin21} J. Hinney, A. S. Prasad, S. Mahmoodian, K. Hammerer, A. Rauschenbeutel, P. Schneeweiss, J. Volz, and M. Schemmer, Unraveling two-photon entanglement via the squeezing spectrum of light traveling through nanofiber-coupled atoms, Phys. Rev. Lett. \textbf{127}, 123602 (2021).

\bibitem{She23} A. S. Sheremet, M. I. Petrov, I. V. Iorsh, A. V. Poshakinskiy, and A. N. Poddubny, Waveguide quantum electrodynamics: Collective radiance and photon-photon correlations, Rev. Mod. Phys. \textbf{95}, 015002 (2023).

\bibitem{Kus23} K. J. Kusmierek, S. Mahmoodian, M. Cordier, J. Hinney, A. Rauschenbeutel, M. Schemmer, P. Schneeweiss, J. Volz, and K. Hammerer, Higher-order mean-field theory of chiral waveguide QED, SciPost Phys. Core \textbf{6}, 041 (2023).

\bibitem{Cor23} M. Cordier, M. Schemmer, P. Schneeweiss, J. Volz, and A. Rauschenbeutel, Tailoring photon statistics with an atom-based two-photon interferometer, Phys. Rev. Lett. \textbf{131}, 183601 (2023).

\bibitem{Mas23} L. Masters, X. X. Hu, M. Cordier, G. Maron, L. Pache, A. Rauschenbeutel, M. Schemmer, and J. Volz, On the simultaneous scattering of two photons by a single two-level atom, Nat. Photonics \textbf{17}, 972 (2023).

\bibitem{scatt} 
 M. Schemmer,   M. Cordier,  L.  Pache,  P. Schneeweiss,   J. Volz, and A. Rauschenbeutel, Simple analytical model describing the collective nonlinear response of an ensemble of two-level emitters weakly coupled to a waveguide (2024), arXiv:2410.21202 [quant-ph].
 





\bibitem{Mol98} A. F. Molisch and B. P. Oehry, \textit{Radiation Trapping in Atomic Vapours} (Oxford University Press, Oxford, 1998).

\bibitem{Fle00} M. Fleischhauer and M. D. Lukin, Dark-state polaritons in electromagnetically induced transparency, Phys. Rev. Lett. \textbf{84}, 5094 (2000).

\bibitem{Fle02} M. Fleischhauer and M. D. Lukin, Quantum memory for photons: Dark-state polaritons, Phys. Rev. A \textbf{65}, 022314 (2002).

\bibitem{Gro82} M. Gross and S. Haroche, Superradiance: An essay on the theory of collective spontaneous emission, Phys. Rep. \textbf{93}, 301 (1982).

\bibitem{Tan11} H. Tanji-Suzuki, I. D. Leroux, M. H. Schleier-Smith, M. Cetina, A. T. Grier, J. Simon, and V. Vuleti\'{c}, Interaction between atomic ensembles and optical resonators: Classical description, in \textit{Advances in Atomic, Molecular, and Optical Physics}, Vol. 60, pp. 201–237 (Academic Press, 2011).

\bibitem{Barr16} D. Barredo, S. de Léséleuc, V. Lienhard, T. Lahaye, and A. Browaeys, An atom-by-atom assembler of defect-free arbitrary two-dimensional atomic arrays, Science \textbf{354}, 1021 (2016).

\bibitem{Vetsch10} E. Vetsch, D. Reitz, G. Sagué, R. Schmidt, S. T. Dawkins, and A. Rauschenbeutel, Optical interface created by laser-cooled atoms trapped in the evanescent field surrounding an optical nanofiber, Phys. Rev. Lett. \textbf{104}, 203603 (2010). 
 
 
	 
	
	
\end{thebibliography}
 \end{document}